\pdfoutput=1

\documentclass[11pt,twoside,a4paper,cmspaper,final,collab]{cms-tdr}
\def\svnVersion{298490}\def\cmsCernNo{2013-037}\def\cmsCernDate{\today}\def\cmsMessage{Published in Physics Letters B as \href{http://dx.doi.org/10.1016/j.physletb.2015.07.037}{\doi{10.1016/j.physletb.2015.07.037.}}}

\begin{document}\cmsNoteHeader{BPH-12-006}

\hyphenation{had-ron-i-za-tion}
\hyphenation{cal-or-i-me-ter}
\hyphenation{de-vices}
\RCS$Revision: 294971 $
\RCS$HeadURL: svn+ssh://svn.cern.ch/reps/tdr2/papers/BPH-12-006/trunk/BPH-12-006.tex $
\RCS$Id: BPH-12-006.tex 294971 2015-07-01 20:55:14Z alverson $
\newlength\cmsFiguresWidth
\ifthenelse{\boolean{cms@external}}{\setlength\cmsFiguresWidth{0.85\columnwidth}}{\setlength\cmsFiguresWidth{0.4\textwidth}}
\ifthenelse{\boolean{cms@external}}{\providecommand{\cmsLeft}{top}}{\providecommand{\cmsLeft}{left}}
\ifthenelse{\boolean{cms@external}}{\providecommand{\cmsRight}{bottom}}{\providecommand{\cmsRight}{right}}
\providecommand{\PgUn}{\ensuremath{\PgU\mathrm{(}n\mathrm{S)}}\xspace}
\ifthenelse{\boolean{cms@external}}{\providecommand{\suppMaterial}{the supplemental material~\cite{suppMaterial}\xspace}} {\providecommand{\suppMaterial}{Appendix~\ref{app:tables}\xspace}}
\ifthenelse{\boolean{cms@external}}{\providecommand{\suppMaterialSecond}{the supplemental material\xspace}} {\providecommand{\suppMaterialSecond}{Appendix~\ref{app:tables}\xspace}}
\ifthenelse{\boolean{cms@external}}{\providecommand{\suppMaterialRef}[2]{#1}}{\providecommand{\suppMaterialRef}[2]{#2}}
\cmsNoteHeader{BPH-12-006}
\title{Measurements of the $\PgUa$, $\PgUb$, and $\PgUc$ differential cross sections in pp collisions at $\sqrt{s} = 7$\TeV}

\date{\today}

\abstract{
Differential cross sections as a function of transverse momentum \pt are presented for the production of $\PgUn$ ($n = 1$, 2, 3) states decaying into a pair of muons. Data corresponding to an integrated luminosity of 4.9\fbinv in pp collisions at $\sqrt{s} = 7$\TeV were collected with the CMS detector at the LHC.  The analysis selects events with dimuon rapidity $\abs{y}< 1.2$ and dimuon transverse momentum in the range $10 < \pt < 100$\GeV.  The measurements show a transition from an exponential to a power-law behavior at $\pt \approx 20$\GeV for the three $\Upsilon$ states.  Above that transition, the $\PgUc$ spectrum is significantly harder than that of the $\PgUa$. The ratios of the $\PgUc$ and $\PgUb$ differential cross sections to the $\PgUa$ cross section show a rise as \pt increases at low $\pt$, then become flatter at higher \pt.
}

\hypersetup{%
pdfauthor={CMS Collaboration},%
pdftitle={Measurements of the Upsilon(1S), Upsilon(2S), and Upsilon(3S) differential cross sections in pp collisions at sqrt(s) = 7 TeV},%
pdfsubject={CMS},%
pdfkeywords={CMS, Upsilon, B-Physics, cross section}}

\maketitle

\section{Introduction}

Hadronic production of S-wave \bbbar mesons has been extensively studied for many years. At the CERN LHC, the CMS~\cite{xs1,cmsx2}, ATLAS~\cite{atlasxs}, and LHCb~\cite{lhcb} Collaborations have published results on \PgUn   ($n$ = 1, 2, 3) production cross sections times dimuon branching fractions in $\Pp\Pp$ collisions at $\sqrt{s} = 7\TeV$ as a function of the $\PgU$ transverse momentum \pt, rapidity $y$, and polarization~\cite{cmspol}. The CMS and ATLAS \pt and $\abs{y}$ distributions in the central rapidity region $\abs{y} < 2.0$ are similar in shape to those from $\Pp\Pap$ production at $\sqrt{s} =1.96$\TeV, as measured by the D0~\cite{d0xs} and CDF~\cite{cdfxs} experiments at the Tevatron. Neither the ATLAS nor the CMS results show any statistically significant rapidity dependence of the cross section in the central region. The CMS analyses cover the \pt range up to 50\GeV, while the ATLAS results go to 70\GeV.

In this Letter we present a measurement of the differential production cross sections of the three lowest-mass \PgUn   states in $\Pp\Pp$ collisions at $\sqrt{s} = 7\TeV$ up to $\pt = 100$\GeV, reaching higher \pt than previous measurements.  We measure the \pt dependence of the \PgUn   differential cross section times the branching fraction to \MM using the 2011 data set, corresponding to an integrated luminosity of 4.9\fbinv. The measured cross sections include feeddown from higher \bbbar excitations.

Measurements of S-wave \bbbar mesons provide an important probe of quantum chromodynamics (QCD). There are several models that predict differential cross section shapes at high \PgUn   \pt in $\Pp\Pp$ collisions.  A common feature of all the models is that different contributing terms have different \pt variations, some of which are power-law forms.  The nonrelativistic QCD (NRQCD) approach~\cite{NRQCD1,NRQCD2} uses an effective field theory to factorize the perturbative term and nonpertubative long-distance matrix element (LDME) terms. A good description of early LHC results for \PgUa \,production for $\pt < 30$\GeV was achieved using NRQCD with next-to-leading-order (NLO) corrections~\cite{wang}.  However, there are theoretical corrections to perturbative NRQCD that have characteristic power-law behavior at high \pt, and measurements at high \pt can help to clarify the theoretical picture~\cite{whitepaper, Bodwin}.   The NLO NRQCD calculation has recently been extended to treat all three \PgUn   states~\cite{gong}.  The updated calculation includes not only NLO terms but also uses LDMEs computed using only high-\pt data.  Color singlet models (CSM) with higher-order \pt-dependent corrections~\cite{CSM} and the \kt-factorization model~\cite{baranov} are consistent with data from the LHC for \pt approaching 50\GeV.  A recent analysis of quarkonium polarization and production measurements found that raising \pt thresholds stabilizes the fits in evaluating the LDMEs~\cite{Faccioli:2014cqa}. At higher \pt different corrections become dominant in these models.  New data at high \pt will challenge all the current approaches.

\section{CMS detector}

The central feature of the CMS apparatus is a superconducting solenoid of 6\unit{m} internal diameter having a 3.8\unit{T} field. Within the superconducting solenoid volume are a silicon pixel and strip tracker, a lead tungstate crystal electromagnetic calorimeter, and a brass and scintillator hadron calorimeter. Muons are measured in gas-ionization detectors embedded in the steel flux-return yoke outside the solenoid, with detection planes are made using three technologies: drift tubes, cathode strip chambers, and resistive-plate chambers.  Muons are measured in the pseudorapidity range $\abs{\eta}< 2.4$.

The silicon tracker measures charged particles within the pseudorapidity range $\abs{\eta}< 2.5$. It consists of 1440 silicon pixel and 15\,148 silicon strip detector modules and provides a typical transverse impact parameter resolution of 25--90\mum.  Matching muons to tracks measured in the silicon tracker results in a transverse momentum resolution between 1\% and 2.8\%, for \pt values up to 100\GeV~\cite{Chatrchyan:2012xi}.

The first level of the CMS trigger system, composed of custom hardware processors, uses information from the calorimeters and muon detectors to select the most interesting events in a fixed time interval of less than 4\mus. The high-level trigger processor farm further decreases the event rate from around 100\unit{kHz} to around 400\unit{Hz}, before data storage. A more detailed description of the CMS detector, together with a definition of the coordinate system used and the relevant kinematic variables, can be found in Ref.~\cite{cmsdet}.

\section{Differential cross section measurement methodology \label{csm}}
Event selection starts with a dimuon trigger involving the silicon tracker and muon systems. The trigger, which is exposed to the full integrated luminosity, requires at least two muons with dimuon rapidity $\abs{y}< 1.25$, dimuon invariant mass $8.5 < M_{\Pgm\Pgm} <  11.5$\GeV, and a dimuon vertex fit with a $\chi^{2}$ probability $>$0.5\%. The trigger selects only pairs of muons that bend away from each other in the magnetic field (``seagull selection''), i.e., events for which the difference in azimuthal angle between the positively charged and negatively charged muons is less than zero. Requiring that muon trajectories do not cross in the transverse plane improves the muon efficiency. Trigger \pt thresholds varied from 5--9\GeV as the beam conditions changed. Offline selection criteria required $\pt>10$\GeV, $\abs{y}< 1.2$, and a dimuon vertex fit $\chi^{2}$ probability $>$1\%.  Standard CMS quality requirements are used to identify muons and muons are restricted to $\abs{\eta(\mu)} < 1.6$. The muon tracks are required to have at least ten hits in the silicon tracker, at least one hit in the silicon pixel detector, and be matched with at least one segment of the muon system. The muon track fit quality must have a $\chi^{2}$ per degree of freedom of less than 1.8. The distance of the track from the closest primary vertex must be less than 15\unit{cm} in the longitudinal direction and 3\unit{cm} in the transverse direction. The following kinematic requirements are also imposed to ensure accurate muon detection efficiency evaluation:
\begin{equation}
\begin{aligned}
 &\pt(\Pgm)>3\GeV&\text{ for }  1.4 <\abs{\eta (\Pgm )}&<1.6,\\
 &\pt(\Pgm)>3.5\GeV& \text{ for }  1.2 <\abs{\eta (\Pgm )}&<1.4,\\
 &\pt(\Pgm)>4.5\GeV& \text{ for }  \abs{\eta (\Pgm )}&<1.2.
\end{aligned}
\label{muon_eta}
\end{equation}

The differential cross sections are measured for two rapidity ranges: $\abs{y} \leq 0.6$ and $0.6<\abs{y}<1.2$, as well as for the entire range $\abs{y}<1.2$.  In each rapidity range the data are binned in \pt, with bin edges at 2\GeV intervals between 10 and 40\GeV, then wider bins with edges at 43, 46, 50, 55, 60, 70, and 100\GeV.

The \PgUn   differential cross section times dimuon branching fraction, integrated over either of the two $\abs{y}$ ranges and in a given \pt bin of width $\Delta \pt$, is
\ifthenelse{\boolean{cms@external}}{
\begin{multline}
\left.\frac{\rd\sigma\Bigl(\Pp\Pp \to \PgUn\Bigr)}{\rd\pt}\right|_{\abs{y}\text{ range}} {\mathcal{B}} \Bigl(\PgUn\to\MM\Bigr)=\\
\frac{ N^{\text{fit}}_{\PgUn} (\pt)}
{ L  \Delta \pt   \epsilon_{\Pgm\Pgm}(\pt)  \mathcal{A}(\pt) \epsilon_{\mathrm{sg}}\epsilon_{\mathrm{vp}}},
\label{diff_xs}
\end{multline}
}{
\begin{equation}
\left.\frac{\rd\sigma\Bigl(\Pp\Pp \to \PgUn\Bigr)}{\rd\pt}\right|_{\abs{y}\text{ range}} {\mathcal{B}} \Bigl(\PgUn\to\MM\Bigr)=
\frac{ N^{\text{fit}}_{\PgUn} (\pt)}
{ L  \Delta \pt   \epsilon_{\Pgm\Pgm}(\pt)  \mathcal{A}(\pt) \epsilon_{\mathrm{sg}}\epsilon_{\mathrm{vp}}},
\label{diff_xs}
\end{equation}
}
where $N^{\text{fit}}_{\PgUn}$ is the fitted number of \PgUn   events from the dimuon invariant mass distribution in a \pt bin for the selected $\abs{y}$ range, $\epsilon_{\Pgm\Pgm}$ is the dimuon efficiency, $L$ is the integrated luminosity, $\mathcal{A}$ is the polarization-corrected acceptance, $\epsilon_{\mathrm{sg}}$ is the efficiency of the seagull selection, and $\epsilon_{\mathrm{vp}}$ is the efficiency of the dimuon vertex $\chi^{2}$ probability requirement. The efficiency and acceptance determinations are described below.

The total yield $N^{\text{fit}}_{\PgUn}$ for the three \PgU($n$S) states in the rapidity range $\abs{y} < $1.2 are $412\,900 \pm 600$ \PgUa\, events,
$151\,700 \pm 400$ \PgUb\, events, and $111\,100 \pm 300$ \PgUc\, events, where the uncertainties are statistical only.  The fine granularity of the CMS tracker kept the
 efficiency independent of changes in the LHC instantaneous luminosity throughout the $\sqrt{s} = 7$\TeV operations.

\subsection{Efficiency factors}

The dimuon efficiency for a given event is parameterized as:
\begin{equation}
 \epsilon_{\Pgm\Pgm} \equiv \epsilon_1[ \pt(\mu_1),\eta(\mu_1)]\;  \epsilon_2[ \pt(\mu_2),\eta(\mu_2)]\; \rho,
\label{dimuon_eff}
\end{equation}
where $\epsilon_i[ \pt(\mu_i),\eta(\mu_i)]$ is the overall single-muon quality and trigger efficiency. The kinematic dependence of the $\rho$ factor was determined in a study based on Monte Carlo (MC) simulation using \EVTGEN ~\cite{evtgen} with a detector simulation performed with \GEANTfour~\cite{GEANT4}. The parameter $\rho$ accounts for the possibility that two genuine muons can be merged during the reconstruction or trigger selection, causing an inefficiency. It was found to depend on the quadrature sum of the differences $\Delta$\PT/(637\GeV), $\Delta \eta$, and $1.2 \Delta \phi$ between the two muons.  The MC simulation result was validated by measuring the $\rho$ factor with \PgU ($nS$) events reconstructed using a data set that required only a single-muon trigger. In events such as those with $\pt < 50$\GeV, where the muons are well separated, $\rho$ = 1. For high-\pt events of $\pt > 80$\GeV, where the muons are closer together, $\rho$ drops to approximately 0.7.

The single-muon efficiencies are measured using the tag-and-probe approach based on control samples in data, as described in Ref.~\cite{tandp}, times the tracking efficiency ($0.99 \pm 0.01$), determined from MC simulation.  We assume that the dimuon efficiency within each \PgUn mass region is the same for signal and background. The dimuon efficiency $\epsilon_{\mu\mu}$ for a given (\pt, $\abs{y}$)  is obtained by averaging the calculated event dimuon efficiency $\epsilon_{\mu\mu}$ for each data event in the bin. This is done separately for the three $\PgU$ states, using a mass range of $\pm$200\MeV for the $\PgUa$ and $\pm$100\MeV for the higher-mass states.  The narrower range for the $\PgUb$ and $\PgUc$ states is chosen because of the closeness in mass of these two states. The average efficiency, $\epsilon_{\Pgm\Pgm}$, is typically 0.75--0.80. For all (\PT,$y$) bins the systematic difference between averaging in  $\epsilon_{\Pgm\Pgm}$ or $1/\epsilon_{\Pgm\Pgm}$  can be neglected in comparison to the quoted systematic uncertainty due to the single muon efficiencies.  To determine $\epsilon_\mathrm{sg}$, we note that there is a 50\% probability that a \PgUn state will decay in the seagull configuration. It was verified in MC simulation that $\epsilon_\mathrm{sg} = 0.5$. The efficiency $\epsilon_{\mathrm{vp}}$ for the dimuon vertex fit $\chi^{2}$ probability requirement is determined to be $0.99 \pm 0.01$ from MC simulation, where the uncertainty is statistical.  This efficiency was validated in data using events from a trigger that did not require vertex selection. We also computed the total acceptance and efficiency product in the MC simulation and compared it with the result based on the factorized approach. The results agreed over the entire \PT range of the measurement.

\subsection{Acceptance}

For each \PgUn   state the acceptance $\mathcal{A}$ is computed in each (\pt, $\abs{y}$) bin and defined as the fraction of its dimuon decays that satisfy the single-muon kinematic selections given by Eq.~(\ref{muon_eta}). The acceptances are computed using generator-level muons, then repeated using reconstructed muons in the full simulation study. The results agree to better than 2\% at all \pt values.
Differences are contained within the systematic uncertainty band (Section 4.3) assigned for the muon reconstruction. To account for the effect of the \PgUn polarization on the muon angular distribution, each MC simulation event is weighted by an angular factor $w$:
\ifthenelse{\boolean{cms@external}}{
\begin{multline}
w=\frac{3}{4\pi}\Biggl(\frac{1}{3+\lambda_{\theta}}\Biggr)\,\times\\\left(1+\lambda_{\theta}\cos^{2}\theta+\lambda_{\phi}\sin^{2}\theta \cos2\phi+\lambda_{\theta\phi}\sin2\theta\cos\phi\right),
\end{multline}
}{
\begin{equation}
w=\frac{3}{4\pi}\Biggl(\frac{1}{3+\lambda_{\theta}}\Biggr)\left(1+\lambda_{\theta}\cos^{2}\theta+\lambda_{\phi}\sin^{2}\theta \cos2\phi+\lambda_{\theta\phi}\sin2\theta\cos\phi\right),
\end{equation}
}
where $\lambda_{\theta},\lambda_{\phi}, \lambda_{\theta\phi}$
are the measured polarization parameters~\cite{cmspol}, $\theta$
is the polar angle, and $\phi$ the azimuthal angle of the positively charged muon in the \PgUn   helicity frame (HX). The polarization was measured in the range $10 < \pt < 50$\GeV in the same two rapidity bins as this analysis. The measured polarization parameters do not show a statistically significant dependence on \pt. We linearly interpolate each of the measured polarization parameters in \pt. Linear interpolation is also used for the 68.3\% confidence level (CL) uncertainties in the polarization measurements to determine the uncertainty in the three parameters from the analysis.  The polarization parameters for $\pt > 50$\GeV are taken to be the average of the measured values for $10 < \pt < 50$\GeV. The largest measured absolute uncertainty for each parameter is used for the extrapolated uncertainties because the spread in nominal values is small.  The acceptance is computed initially using a flat \pt distribution within a bin, then reweighted after fitting the measured \pt distribution to a functional form (see Section 5). The acceptances in each \pt bin for the three rapidity intervals are given in \suppMaterial (Tables \suppMaterialRef{7--15}{\ref{AccSumy01S}--\ref{AccSumy23S}}) for the measured polarization central value and the 68.3\% CL uncertainties on the parameters~\cite{cmspol}. In addition, we report the acceptance computed for the hypotheses of zero, 100\% transverse, and 100\% longitudinal polarization that correspond to the parameter values $\lambda_{\phi}$ = $\lambda_{\theta\phi}$ = 0 and $\lambda_{\theta}$ = 0, $+1$, and $-1$ respectively. Because of the agreement in the acceptance when computed with generator-level and reconstructed muons, the cross section results reported here can be scaled to accommodate any other polarization by using a generator-level MC simulation with a given polarization.

\section{Yield determination procedure}
\subsection{Lineshape determination}
The \PgUn   lineshape is determined using the measured muon momenta and their uncertainties, along with a generator-level simulated invariant mass (SIM) distribution including final-state radiation (FSR) effects.  For events in a given (\pt, $\abs{y}$) bin, the distribution of the dimuon invariant mass uncertainty $\zeta$ is computed from the muon track error matrices.

In order to describe the \PgUn   SIM distribution without detector resolution effects, we simulate dimuon events for a given \PgUn state using \EVTGEN and compute the FSR using \PHOTOS ~\cite{PHOTOS1,PHOTOSv2}. The standard \PHOTOS minimum photon energy for the \PgUn   states is $\approx$50 \MeV, which is of the same order as our dimuon invariant mass uncertainty. To improve the description, we extend the photon energy spectrum down to 2 \MeV using a fit of the SIM distribution to the QED inner-bremsstrahlung formula~\cite{PHOTOSv2}.  The systematic uncertainties of the soft photon approximation in \PHOTOS compared to exact QED calculations are discussed in Ref.~\cite{PHOTOSv2}.  For the range of photon energies expected in \PgUn   decays the systematic uncertainty is negligible.

In each rapidity range, the \PgUn   lineshape for a given \pt bin is expressed by a probability density function (PDF) for the signal dimuon mass $M_{\Pgm\Pgm}$.  This function $\mathcal{F}(M_{\Pgm\Pgm}; \,c_w,\,\delta m)$ is the average of $N$ values of the dimuon mass $m_i$ smeared with a resolution $\zeta_i$:
 \begin{equation}
\mathcal{F}(M_{\Pgm\Pgm};\,c_w,\,\delta m) = \frac{1}{N} \sum_{i=1}^N \frac{1}{\sqrt{2 \pi} c_w \zeta_i} \re^{-(M_{\Pgm\Pgm} - m_i - \delta m)^2/2 c_w^2 \zeta_i^2}.
\label{line-shape}
\end{equation}
Each \PgUn   state is handled in the same fashion.  Values of $m_{i}$ and $\zeta_{i}$ are selected by randomly sampling the radiative mass function and the $\zeta$ distribution for that (\pt, $\abs{y}$) bin. Two correction factors are common to all three \PgUn   peaks in a given (\pt, $\abs{y}$) bin: a width scale factor $c_w$, to correct for any $\zeta$ scale difference between data and the MC simulation, and a mass-shift $\delta$m, to correct for any difference in \pt scale between data and the MC simulation.  We sample $N = 25\,000$ ($m_{i},\,\zeta_{i}$) points per \pt bin, stored in a histogram with 0.25\MeV bins to smooth the fluctuations and retain shape features.  This histogram gives the normalized, resolution-smeared mass PDF for a given \PgUn   state in a particular (\pt, $\abs{y}$) bin.  The procedure was validated in MC simulation by generating the lineshape using a subset of generated \PgUa \ events, then fitting the rest of the events with that lineshape. The fitted number of events was consistent with the generated number.

\subsection{Fitting for yields}

To determine the yields of the three states in each \pt and $\abs{y}$ range requires a fit to the dimuon mass distribution in every (\pt, $\abs{y}$) bin.  The total PDF for $M_{\Pgm\Pgm}$ describes the signal and background contributions to the dimuon invariant mass distribution using a signal PDF as defined in Eq.~(\ref{line-shape}) for each of the \PgUn   states, plus a background function. Four background functions are studied: an exponential and a Chebyshev series with maximum order of 0, 1, or 2.

We measure the yield by performing an extended maximum-likelihood fit using \textsc{RooFit}~\cite{Verkerke:2003ir} to determine the number of signal events associated with each normalized signal PDF. To allow cancellation of some common uncertainties in the muon acceptance and efficiency calculation in the measurement of the ratios of $\PgUb$ and $\PgUc$  differential cross sections to that of the $\PgUa$, we perform an additional fit normalized to the \PgUa \ yield. For each \pt bin the optimal background function is determined using the Akaike Information Criterion (AIC)~\cite{aic}, taking the function with the largest relative probability, as discussed in Ref.~\cite{aic1}. This method is similar to a maximum-likelihood evaluation, but it adds a term equal to twice the number of free parameters in the fit, thus penalizing addition of free parameters. The  parameters $c_w$ and $\delta m$ are determined from the fit for each \pt bin. Typical values and corresponding uncertainties for $c_w$ and $\delta m$ are $1.04 \pm 0.01$ and $3 \pm 1$\MeV, respectively. The fit correlation matrix shows that their influence on the yields is a small fraction of the statistical uncertainty in each yield.

\begin{figure}[htb!]
\centering
   \includegraphics[width=.45\textwidth]{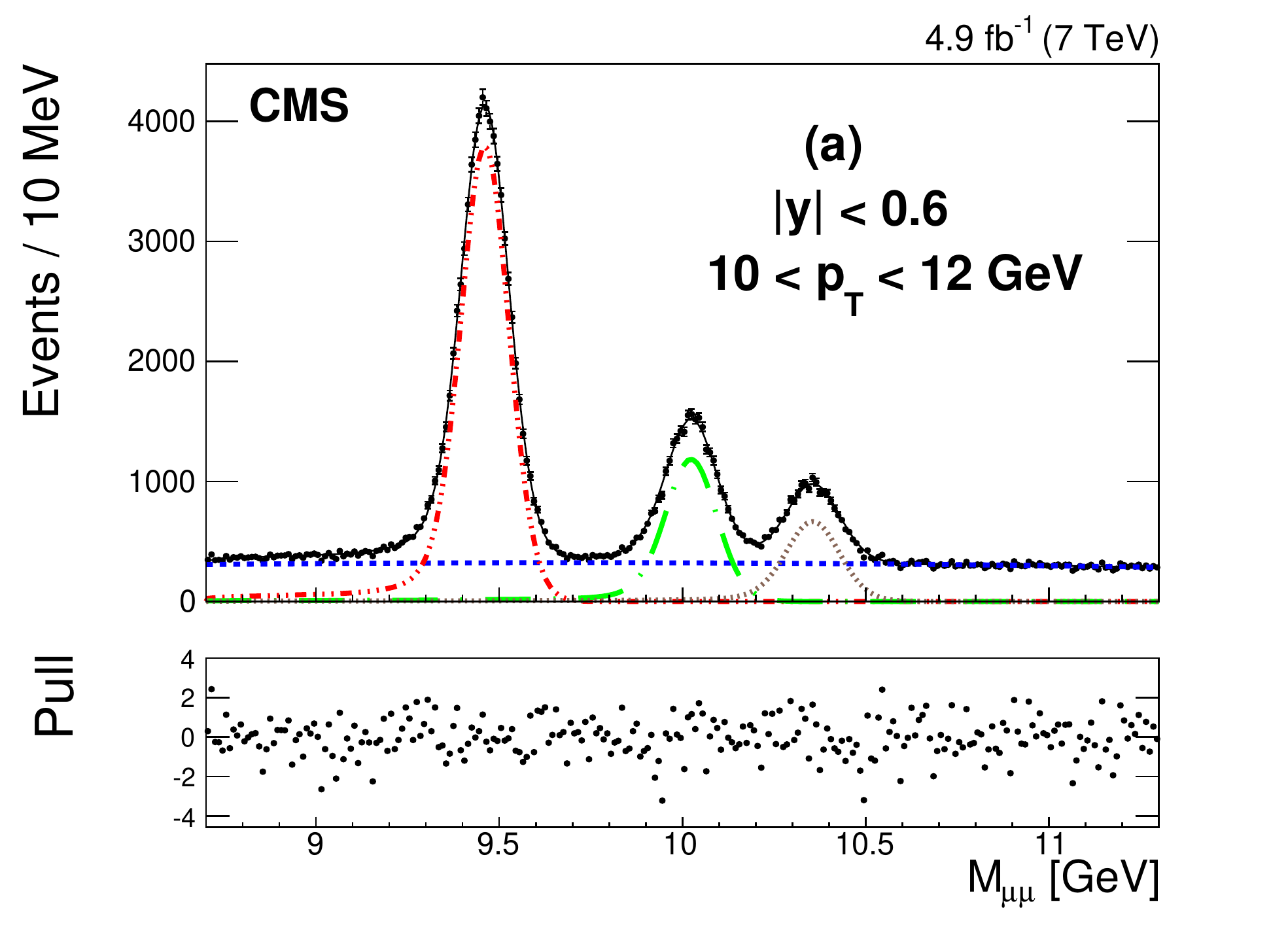}
   \includegraphics[width=.45\textwidth]{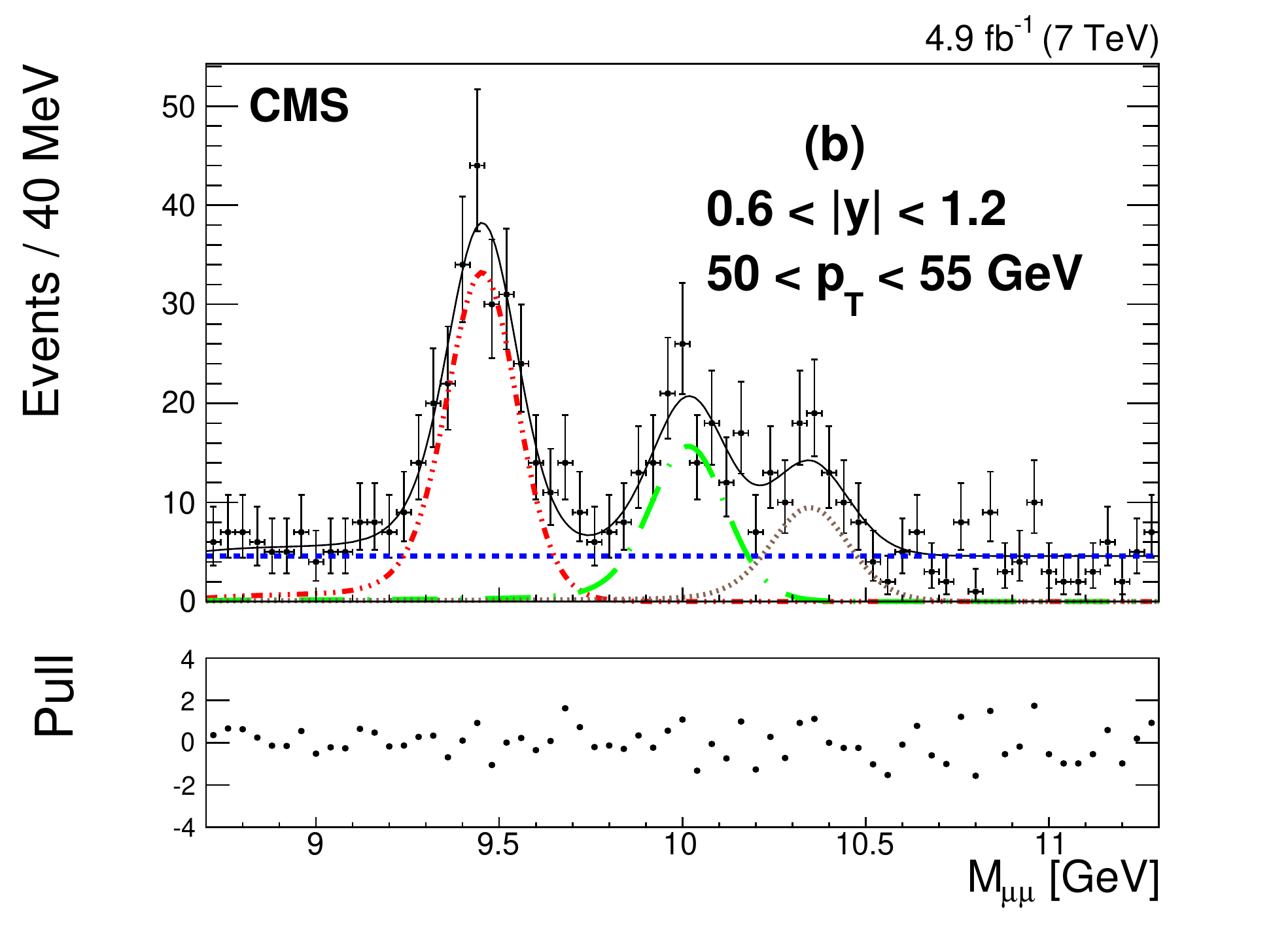}
\caption{ Results of the fits to the dimuon invariant mass distribution for events in two bins: (a): $\abs{y}< 0.6$, $10 < \pt < 12$\GeV and (b): $0.6 < \abs{y}< 1.2$, $50 < \pt < 55$\GeV.  The solid line is the result of the full fit.  The dash-dotted line is the $\PgUa$ signal fit, the long-dashed line is the $\PgUb$ signal fit, and the dotted line is the $\PgUc$ signal fit.  The short-dashed line is the background contribution.  The lower plots show the pull for each mass bin.  \label{binnedmass}}
\end {figure}

The plots in Fig.~\ref{binnedmass} show two examples of fitting the dimuon invariant mass distribution using the lineshape method. The lower plots show the pull, $(N_{\text{data}} - N_{\text{fit}})/\sigma_{\text{data}}$, in each dimuon mass bin, where $N_{\text{data}}$ is the observed number of events in the bin, $N_{\text{fit}}$ is the integral of the fitted signal and background function in that bin, and the uncertainty $\sigma_{\text{data}}$ is the Poisson statistical uncertainty.  As can be seen in Fig.~\ref{binnedmass}, the lineshape description represents the data well, even at high \pt and large rapidity.

\subsection{Systematic uncertainties}
The overall systematic uncertainty in the cross section for a given (\pt, $\abs{y}$) bin includes uncertainties from the background fit method, the lineshape determination, the dimuon efficiency, the acceptance variations due to varying the polarization parameters within their 68.3\% CL ranges, and the integrated luminosity. The systematic uncertainty from the background function is estimated using the maximum difference in yields among background functions with an AIC probability above 5\%~\cite{aic, aic1} relative to the best background choice. An upper limit of 1\% on the systematic uncertainty from the lineshape function determination for all three \PgUn states and all (\pt, $\abs{y}$) bins is estimated by varying the width of the mass region in which the mass resolution parameter $\zeta$ is determined. The efficiency systematic is evaluated by modifying $\epsilon_{\Pgm\Pgm}$ event by event, using the $\pm$1 standard deviation values from the tag-and-probe measurements~\cite{cmspol}.  There is a 1\% systematic uncertainty to account for small variations in $\epsilon_{\Pgm\Pgm}$ as a function of $M_{\Pgm\Pgm}$ observed in the data.   The measured $\rho$ factor values from the experimental determination and from MC simulation agree over the full \pt range.  We assign a systematic uncertainty for $\rho$ of 0.5--5\%, which equals the full difference between the MC simulation and the experimental measurement.  We compute the acceptance systematic uncertainty by raising and lowering all three polarization parameters by their interpolated 68.3\% CL values from Ref.~\cite{cmspol}. The resulting 5--8\% change in the acceptance is used as the systematic uncertainty in the acceptance as tabulated in \suppMaterialSecond (Tables \suppMaterialRef{7--15}{\ref{AccSumy01S}--\ref{AccSumy23S}}). The total systematic uncertainty is found from the quadrature sum of the individual systematic uncertainties. It is comparable to or smaller than the statistical uncertainty for $\pt > 40$\GeV. There is a 2.2\% uncertainty~\cite{lumi} from the integrated luminosity determination that applies to all \pt bins. This uncertainty is not included in the uncertainties displayed in the figures or given in the tables.

\section{Results}
\begin{figure}[h!tb]
\centering
\includegraphics[width=0.48\textwidth]{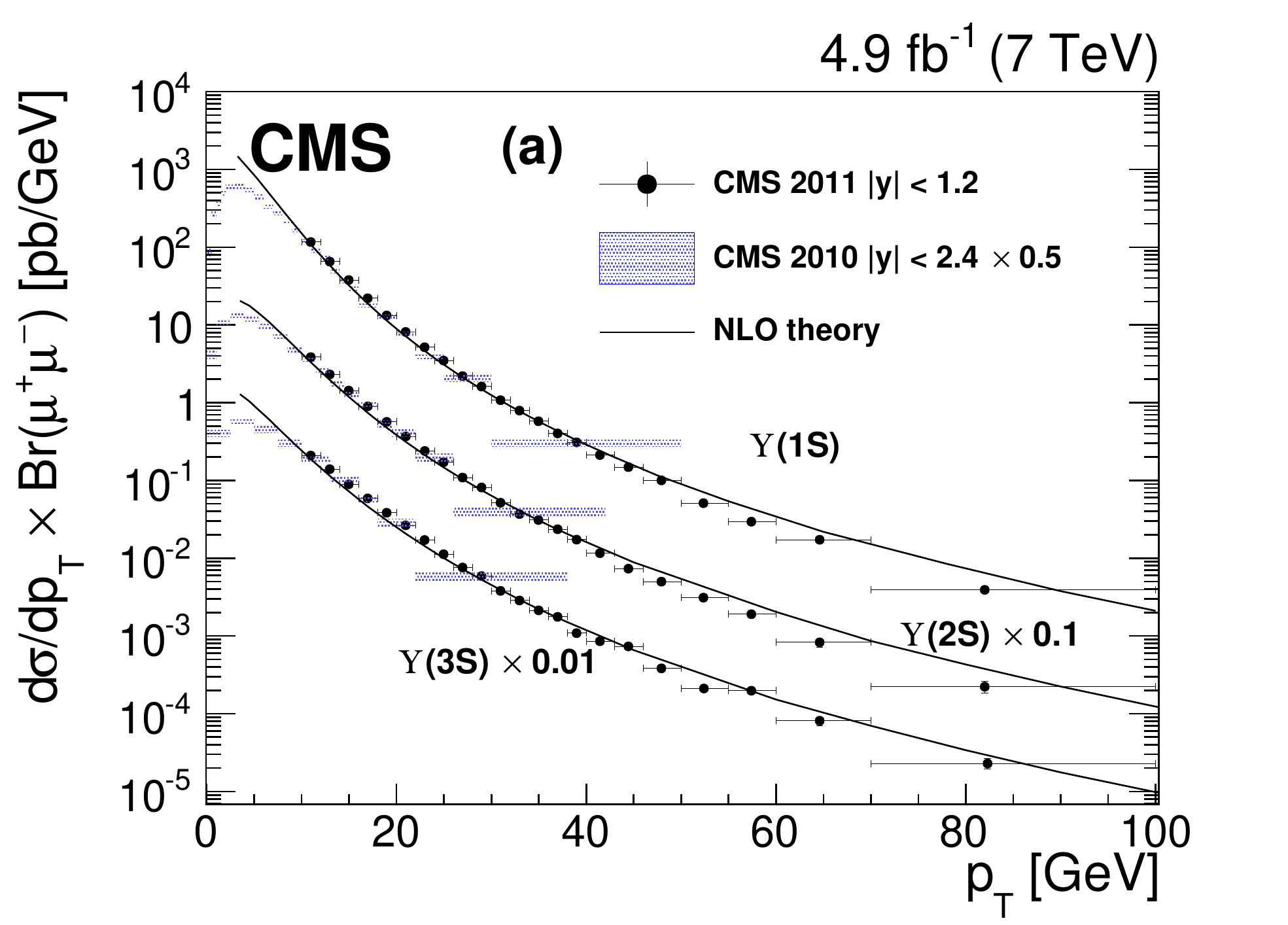}
\includegraphics[width=0.48\textwidth]{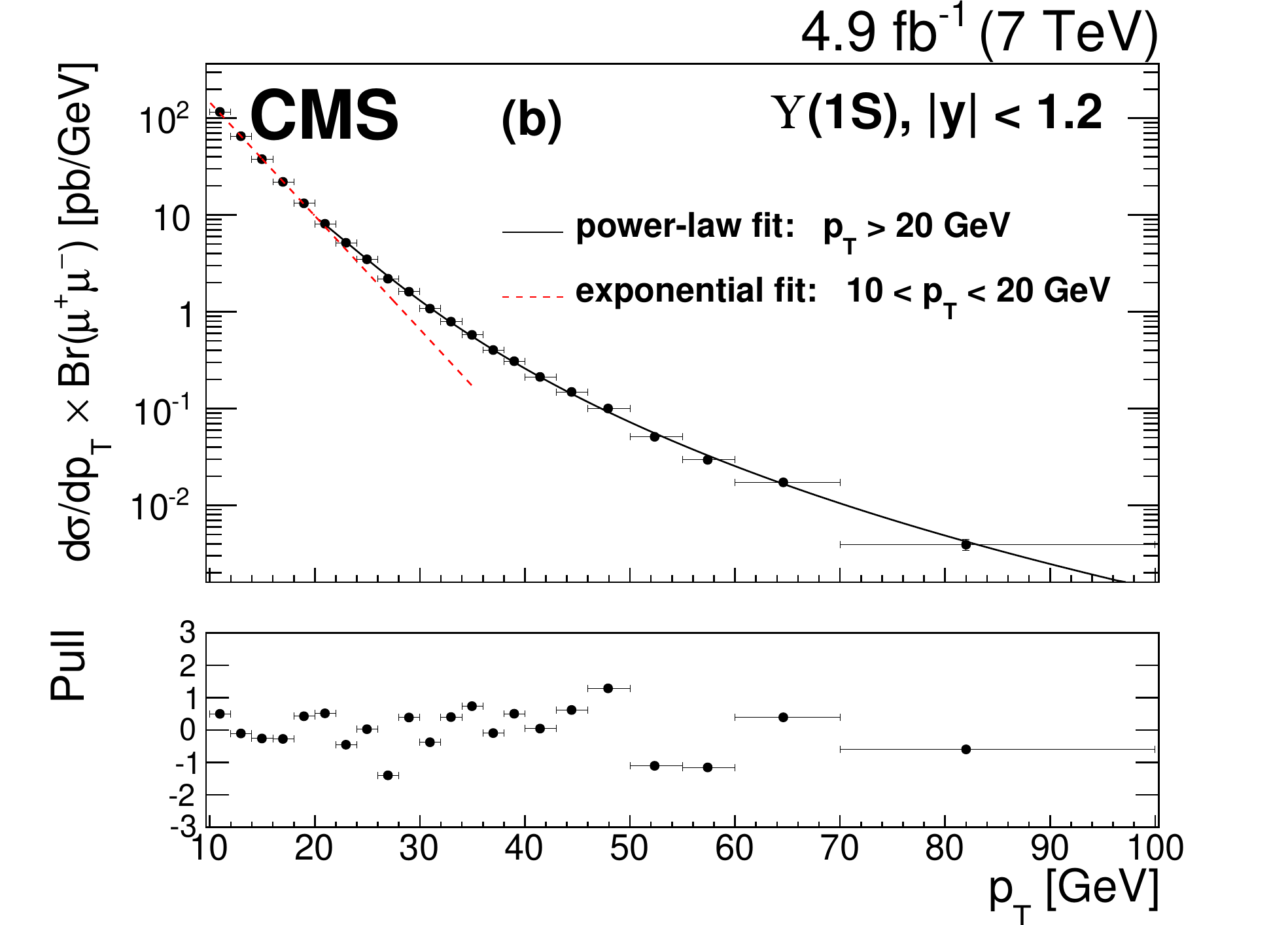}
\caption{(a) The \PgUn differential \pt cross sections times dimuon branching fractions for $\abs{y} < $ 1.2. The $\PgUb$ and $\PgUc$ measurements are scaled by 0.1 and 0.01, respectively, for display purposes.  The vertical bars show the total uncertainty, excluding the systematic uncertainty in the integrated luminosity. The horizontal bars show the bin widths. Previous CMS measurements for $\abs{y} < $ 2.4 are shown as cross-hatched areas~\cite{cmsx2}. These results have been scaled by 0.5 to account for the smaller $\abs{y}$ range in the latest measurement, where the scaling assumes that the rapidity distribution is flat. The solid lines are the NLO calculations from Ref.~\cite{gong} extended by the authors to cover the range $\pt < 100$\GeV. (b) Details of the parametrized cross section fit described in the text for \PgUa \, with $\abs{y} < 1.2$.  In this plot the solid line is the result of the power-law fit (see Eq.~\ref{phenom}) for $\pt > 20$\GeV.  The dashed line shows an exponential fit to the data for $10 < \pt < 20$\GeV.  The lower plot shows the pulls of the fit as defined in the text. \label{dsdpt}}
\end{figure}

The measured \PgUn   differential cross sections versus \pt are shown in Fig.~\ref{dsdpt} over the full rapidity range $\abs{y} < 1.2$.  The vertical bars on the points in Fig.~\ref{dsdpt} show the statistical and systematic uncertainties added in quadrature.  Earlier CMS measurements~\cite{cmsx2} are shown for comparison, scaled by 0.5 to account for the smaller $\abs{y}$ range in the latest measurement, where the scaling assumes that the rapidity distribution is flat. The  \PgUn   differential cross sections peak near $\pt = 4$\GeV, as seen in Fig.~\ref{dsdpt}. Their shape can be described by an exponential function for $10 \lesssim \pt \lesssim 20$\GeV, while for $\pt \gtrsim20$\GeV the data lie above the exponential and the slope changes. Therefore, we fit the high-\pt measurements for each \PgUn state using a power-law parametrization:
\ifthenelse{\boolean{cms@external}}{
\begin{multline}
\left.\frac{\rd\sigma\Bigl(\Pp\Pp \to \PgUn\Bigr)}{\rd\pt}\right|_{\abs{y} \text{ range}} { \mathcal{B}} \Bigl(\PgUn\to\MM\Bigr)=\\ \frac{A}{C + \Bigl(\frac{\pt}{p_0}\Bigr)^{\alpha}},
\label{phenom}
\end{multline}
}{
\begin{equation}
\left.\frac{\rd\sigma\Bigl(\Pp\Pp \to \PgUn\Bigr)}{\rd\pt}\right|_{\abs{y} \text{ range}} { \mathcal{B}} \Bigl(\PgUn\to\MM\Bigr)= \frac{A}{C + \Bigl(\frac{\pt}{p_0}\Bigr)^{\alpha}},
\label{phenom}
\end{equation}
}
where $A$ is a normalization with units of pb/\GeVns. The value of $p_0$ is fixed to 20\GeV and has no influence on the exponent $\alpha$, which describes the curvature of the function.  The differential cross section fits are evaluated using the integral value of the function over the \pt range of each bin, and the results are given in Table~\ref{fit_par}. The bin centers are determined by the functional-weight method described in~\cite{Lafferty}, using the exponential fit for $\pt < 20$\GeV and the power-law form in Eq.~\ref{phenom} for $\pt > 20$\GeV. Shifts from the \pt-weighted mean values are negligible in all except the highest-\pt bin, where using the functional weight moves the bin center from 79 to 82\GeV.  Tables~\suppMaterialRef{1--3}{\ref{xs_table_y0}--\ref{xs_table_y2}} in \suppMaterialSecond give the measured values shown in Fig.~\ref{dsdpt} as well as for the two rapidity ranges $\abs{y} <0.6$ and 0.6 $< \abs{y} < 1.2$.

To illustrate the quality of this functional description, Fig.~\ref{dsdpt}~(b) shows the fit results for the $\PgUa$ state with $\abs{y} < 1.2$.  The solid line is the power-law fit for $\pt > 20$\GeV. The dashed line is the exponential fit for $10 < \pt < 20$\GeV.  The lower plot shows, for each \pt bin, the pull determined from the differential cross section value in a (\pt, $\abs{y}$) bin and its total uncertainty.
\begin{table}[ht]
\centering
\topcaption{ \ The values of the parameters in Eq.~(\ref{phenom}) from the power-law fit to $\PgUa$ events with $\pt > 20\GeV$ and  $\abs{y} < $ 1.2, along with the $\chi^2$ value and the number of degrees of freedom $n_{d}$. \label{fit_par}}

\begin{tabular}{ccccc}\\
 \hline
 & $\PgUa$ & $\PgUb$ & $\PgUc$ \\ \hline
A & 14.00 $\pm$ 0.75 & 6.88 $\pm$ 0.48 & 4.01 $\pm$ 0.30 & \\
$\alpha$ & 5.75 $\pm$ 0.07 & 5.62 $\pm$ 0.10 & 5.26 $\pm$ 0.10 & \\
C & 0.45 $\pm$ 0.13 & 0.62 $\pm$ 0.18 & 0.26 $\pm$ 0.15 & \\
$\chi^{2}$ & 8.7 & 11 & 15 & \\
 $n_{d}$ & 14 & 14 & 14 & \\ \hline
\end{tabular}
\end{table}

\begin{figure}[bh!t]
\centering
\includegraphics[width=0.48\textwidth]{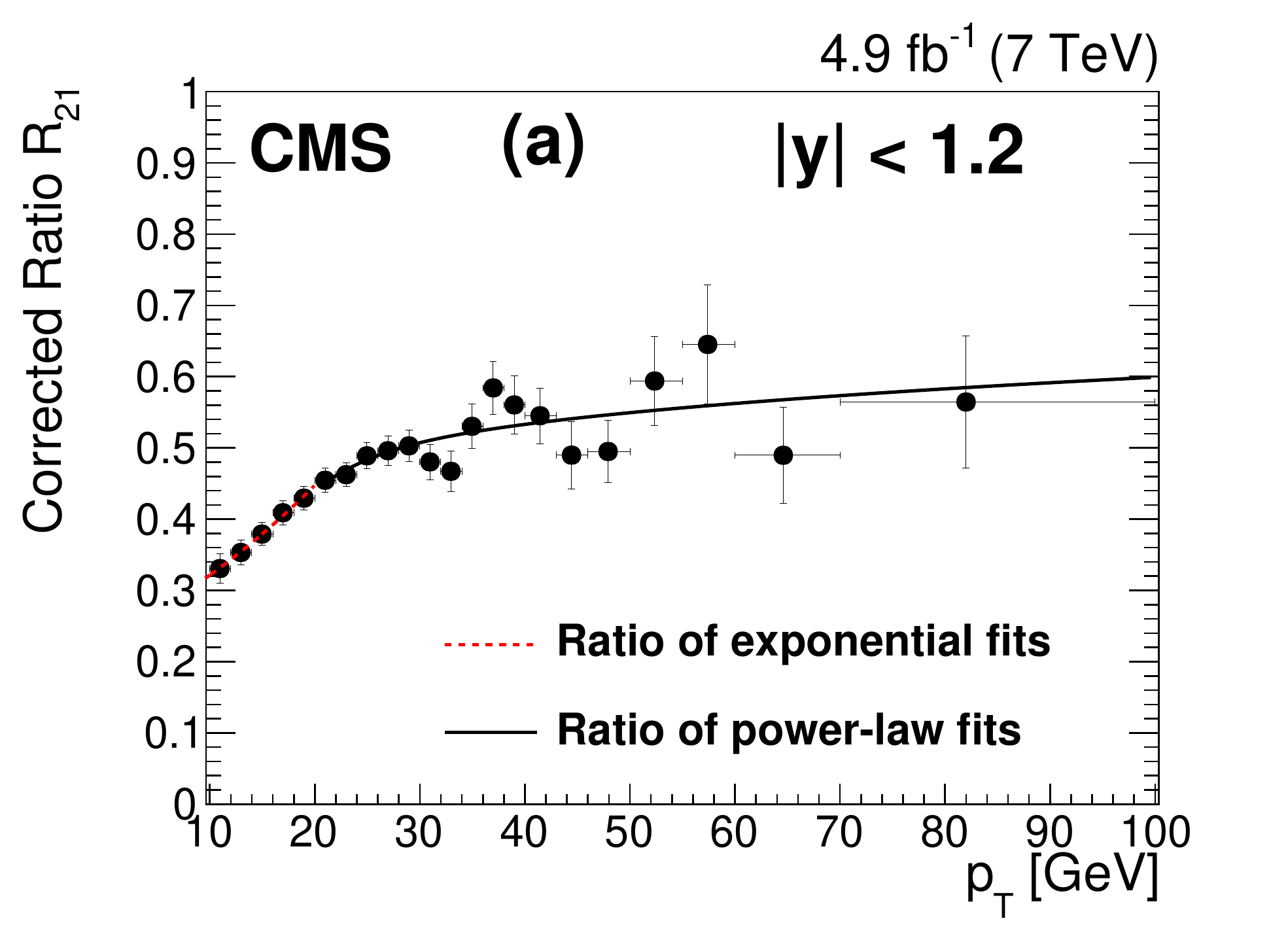}
\includegraphics[width=0.48\textwidth]{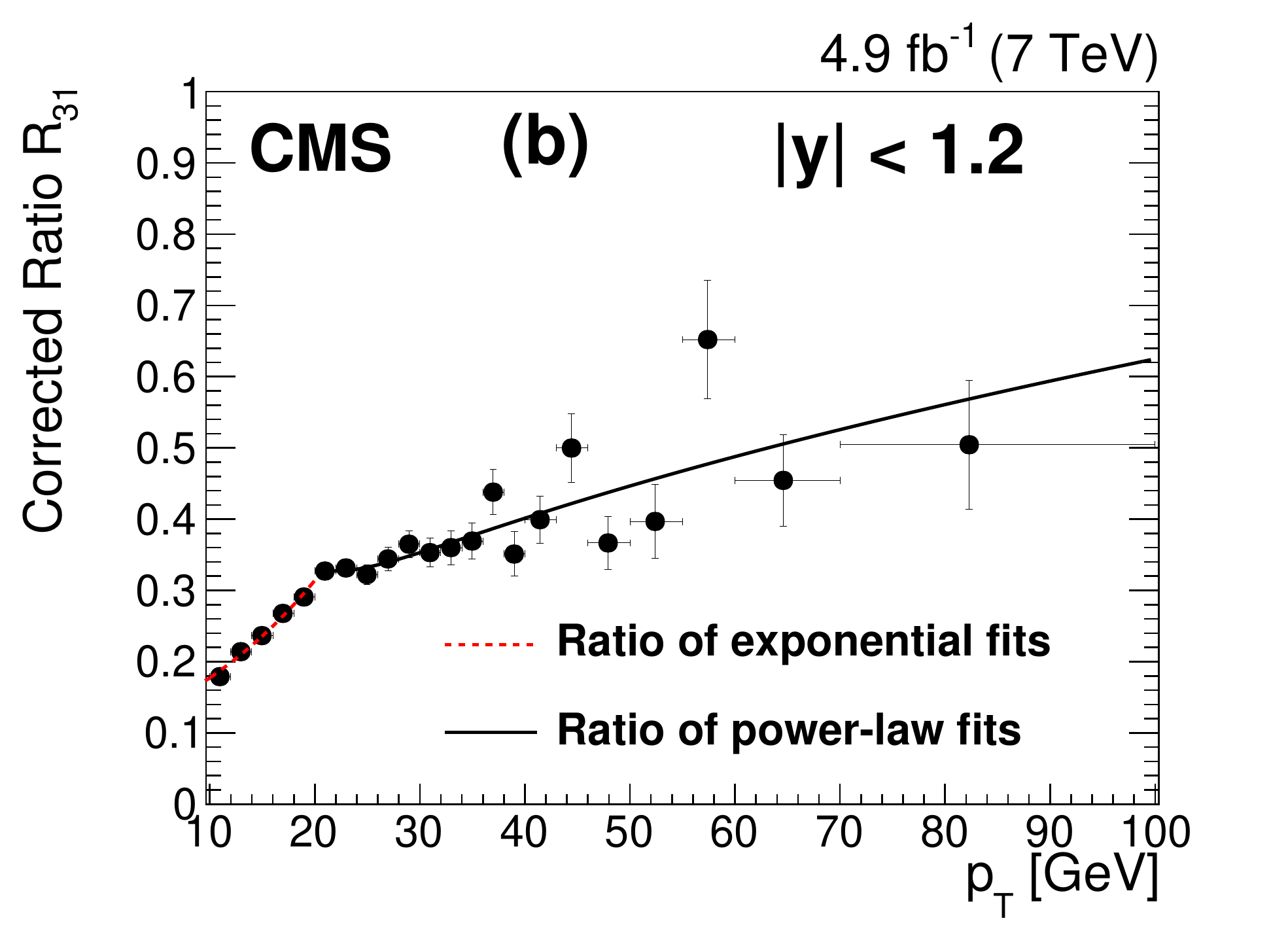}
\caption{Measured differential cross section ratios as a function of \pt. Corrected yield ratios: (a) $R_{21}$;   (b) $R_{31}$.  The dashed line is the ratio of the exponential fits to the individual differential cross sections for $10 < \pt <20$\GeV.  The solid line is the ratio of the corresponding power-law fits for $\pt > 20$\GeV.\label{rat}}
\end{figure}

Next, we consider the \pt dependence of the ratios of the \PgUn   production cross sections times their dimuon branching fractions. The yield fits are redone to compute explicitly the yield ratio $r_{21}$ for $\PgUb$ to $\PgUa$ and $r_{31}$ for $\PgUc$ to \PgUa.  The efficiency ratio is computed for each (\pt, $\abs{y}$) bin.  The polarization-weighted acceptance and its uncertainty is computed for each state separately, and the uncertainties are added in quadrature to determine the uncertainty in the ratio. The corrected yield ratios are $R_{n1}$(\pt, $\abs{y}$) = $r_{n1}$(\pt, $\abs{y}$) $(\mathcal{A}_1  \epsilon_{1})/(\mathcal{A}_n \epsilon_{n})$, where $n$ = 2, 3. The measured corrected ratios are shown in Fig.~\ref{rat} and given in \suppMaterialSecond (Tables~\suppMaterialRef{4--6}{\ref{RatioTabley0}--\ref{RatioTabley2}}).  The rapid rise of both ratios for $\pt < 20$\GeV slows significantly for $\pt \gtrsim 20$\GeV. The curves on the ratio plots are the ratios of the corresponding fitted functions from the individual \PgUn    differential cross section fits (exponential for $\pt< 20$\GeV, power-law for $\pt>20$\GeV). The curves confirm that the change in ratios occurs in the same \pt range in which $\rd\sigma/\rd\pt$ also changes behavior.

The measurements for the ratio $R_{31}$ in Fig.~\ref{rat}(b), found in the supplementary material, can be fit to a linear function and to a constant in order to quantify the visual evidence that the \PgUc \ production is harder than that of the \PgUa.  The linear fit to measurements with $\PT >20\GeV$ has $\chi^2$ probability 0.22, while the fit to a constant has $\chi^2$ probability 2.6 $\times 10^{-5}$.  Thus, with relative probability 85000:1, we can say that $\PgUc$ production is harder than that of the $\PgUa$.  The $\PgUb/\PgUa$ production ratio versus \PT has a similar trend, but the statistical uncertainties are too large to make a definite statement.

\section{Discussion}
Theoretical predictions for the \PgUn   differential cross sections have been previously compared to the first LHC cross section measurements~\cite{wang,CSM,baranov}.  A more recent CMS measurement~\cite{cmsx2} included the currently available predictions from the CSM~\cite{CSM}, valid for $\pt < 35$\GeV, and an unpublished NRQCD prediction that covers the range $\pt < 30$\GeV. The NRQCD + NLO analysis from Ref.~\cite{gong} describes \PgUn   production at Tevatron and LHC energies for $\pt < 50$\GeV.  An extension of these predictions to $\pt = 100$\GeV is compared to the CMS measurements in Fig.~\ref{dsdpt}~(a). The calculations describe the trends of the data for all three \PgUn  states.

The color evaporation model (CEM), a variant of the CSM, predicts that above a minimum $\pt \approx M_{\PgUa}$, all bottomonium states should have the same \pt dependence~\cite{cem}. The measured ratios of the differential cross sections as a function of \pt in Fig.~\ref{rat} show that this is not the case for \pt less than about 40\GeV.

Changing the \PgUn \pt threshold for the data used in calculating the NRQCD predictions results in different LDMEs~\cite{buten1,buten2,wang}.  Recent theoretical work~\cite{Bodwin,Faccioli:2014cqa} has demonstrated the impact of varying the \pt thresholds in NRQCD analyses to study different production amplitude behavior.  These new CMS data provide a significant extension of the \pt range that can be used in evaluating matrix elements and studying \pt-dependent corrections in NRQCD and other models.  The new results on $\PgUc$ production are sufficiently accurate to allow one to focus model building of the \pt behavior on that state, for which feeddown contributions come only from the $\chi_{b}$(3P).

\section{Summary}
Measurements of the differential production cross sections as a function of \pt for the \PgUa, \PgUb, and $\PgUc$ states in $\Pp\Pp$ collisions at $\sqrt{s} = 7\TeV$ have been presented, based on a data sample corresponding to an integrated luminosity of 4.9\fbinv collected by the CMS experiment at the LHC. Not only do these measurements significantly improve the precision of the results in previously analyzed \pt ranges~\cite{xs1,cmsx2,atlasxs}, they also extend the maximum \pt range from 70 to 100\GeV. Evidence has been presented for the first time of the power-law nature of the \pt distributions for all three \PgUn   states at high \pt.  Combined with the CMS \PgUn   polarization results~\cite{cmspol}, the new bottomonium measurements are a formidable challenge to our theoretical understanding of the production of heavy-quark bound states.

\begin{acknowledgments}
We congratulate our colleagues in the CERN accelerator departments for the excellent performance of the LHC and thank the technical and administrative staffs at CERN and at other CMS institutes for their contributions to the success of the CMS effort. In addition, we gratefully acknowledge the computing centers and personnel of the Worldwide LHC Computing Grid for delivering so effectively the computing infrastructure essential to our analyses. Finally, we acknowledge the enduring support for the construction and operation of the LHC and the CMS detector provided by the following funding agencies: BMWFW and FWF (Austria); FNRS and FWO (Belgium); CNPq, CAPES, FAPERJ, and FAPESP (Brazil); MES (Bulgaria); CERN; CAS, MoST, and NSFC (China); COLCIENCIAS (Colombia); MSES and CSF (Croatia); RPF (Cyprus); MoER, ERC IUT and ERDF (Estonia); Academy of Finland, MEC, and HIP (Finland); CEA and CNRS/IN2P3 (France); BMBF, DFG, and HGF (Germany); GSRT (Greece); OTKA and NIH (Hungary); DAE and DST (India); IPM (Iran); SFI (Ireland); INFN (Italy); MSIP and NRF (Republic of Korea); LAS (Lithuania); MOE and UM (Malaysia); CINVESTAV, CONACYT, SEP, and UASLP-FAI (Mexico); MBIE (New Zealand); PAEC (Pakistan); MSHE and NSC (Poland); FCT (Portugal); JINR (Dubna); MON, RosAtom, RAS and RFBR (Russia); MESTD (Serbia); SEIDI and CPAN (Spain); Swiss Funding Agencies (Switzerland); MST (Taipei); ThEPCenter, IPST, STAR and NSTDA (Thailand); TUBITAK and TAEK (Turkey); NASU and SFFR (Ukraine); STFC (United Kingdom); DOE and NSF (USA).

Individuals have received support from the Marie-Curie program and the European Research Council and EPLANET (European Union); the Leventis Foundation; the A. P. Sloan Foundation; the Alexander von Humboldt Foundation; the Belgian Federal Science Policy Office; the Fonds pour la Formation \`a la Recherche dans l'Industrie et dans l'Agriculture (FRIA-Belgium); the Agentschap voor Innovatie door Wetenschap en Technologie (IWT-Belgium); the Ministry of Education, Youth and Sports (MEYS) of the Czech Republic; the Council of Science and Industrial Research, India; the HOMING PLUS program of Foundation for Polish Science, cofinanced from European Union, Regional Development Fund; the Compagnia di San Paolo (Torino); the Consorzio per la Fisica (Trieste); MIUR project 20108T4XTM (Italy); the Thalis and Aristeia programs cofinanced by EU-ESF and the Greek NSRF; and the National Priorities Research Program by Qatar National Research Fund.
\end{acknowledgments}

\bibliography{auto_generated}

\providecommand{\href}[2]{#2}\begingroup\raggedright\begin{thebibliography}{10}%
\makeatletter
\providecommand{\hrefCMSnoop }[0]{\@secondoftwo}%
\makeatother
\providecommand{\doi}{\texttt{doi:}\begingroup \urlstyle{tt}\Url}

\bibitem{xs1}
\hrefCMSnoop {}{{CMS} Collaboration, ``{Measurement of the inclusive $\Upsilon$
  production cross section in pp collisions at $\sqrt{s}=7$ TeV}'',} \textit{
  Phys. Rev. D} \textbf{ 83} (2011) 112004,
  \href{http://dx.doi.org/10.1103/PhysRevD.83.112004}{\doi{10.1103/PhysRevD.83.112004}},
\href{http://www.arXiv.org/abs/1012.5545}{\texttt{arXiv:1012.5545}}.

\bibitem{cmsx2}
\hrefCMSnoop {}{{CMS} Collaboration, ``{Measurement of the $\Upsilon(1S),
  \Upsilon(2S)$, and $\Upsilon(3S)$ cross sections in pp collisions at
  $\sqrt{s}$ = 7 TeV}'',} \textit{ Phys. Lett. B} \textbf{ 727} (2013) 101,
  \href{http://dx.doi.org/10.1016/j.physletb.2013.10.033}{\doi{10.1016/j.physletb.2013.10.033}},
\href{http://www.arXiv.org/abs/1303.5900}{\texttt{arXiv:1303.5900}}.

\bibitem{atlasxs}
\hrefCMSnoop {}{{ATLAS} Collaboration, ``{Measurement of upsilon production in
  7 TeV $pp$ collisions at ATLAS}'',} \textit{ Phys. Rev. D} \textbf{ 87}
  (2013) 052004,
  \href{http://dx.doi.org/10.1103/PhysRevD.87.052004}{\doi{10.1103/PhysRevD.87.052004}},
\href{http://www.arXiv.org/abs/1211.7255}{\texttt{arXiv:1211.7255}}.

\bibitem{lhcb}
\hrefCMSnoop {}{{LHCb} Collaboration, ``{Measurement of $\Upsilon$ production
  in $pp$ collisions at $\sqrt{s}$ = 7 TeV}'',} \textit{ Eur. Phys. J. C}
  \textbf{ 72} (2012) 2025,
  \href{http://dx.doi.org/10.1140/epjc/s10052-012-2025-y}{\doi{10.1140/epjc/s10052-012-2025-y}},
\href{http://www.arXiv.org/abs/1202.6579}{\texttt{arXiv:1202.6579}}.

\bibitem{cmspol}
\hrefCMSnoop {}{{CMS} Collaboration, ``Measurement of the {Y(1S), Y(2S) and
  Y(3S}) Polarizations in pp Collisions at $\sqrt{s}=7$ {TeV}'',} \textit{
  Phys. Rev. Lett.} \textbf{ 110} (2013) 081802,
  \href{http://dx.doi.org/10.1103/PhysRevLett.110.081802}{\doi{10.1103/PhysRevLett.110.081802}},
\href{http://www.arXiv.org/abs/1209.2922}{\texttt{arXiv:1209.2922}}.

\bibitem{d0xs}
\hrefCMSnoop {}{{D0} Collaboration, ``Measurement of Inclusive Differential
  Cross Sections for {$\Upsilon(1S)$} Production in p$\bar{\rm{p}}$ Collisions
  at $\sqrt{s}$ = 1.96 {TeV}'',} \textit{ Phys. Rev. Lett.} \textbf{ 94} (2005)
  232001,
  \href{http://dx.doi.org/10.1103/PhysRevLett.94.232001}{\doi{10.1103/PhysRevLett.94.232001}},
  \href{http://www.arXiv.org/abs/hep-ex/0502030}{\texttt{arXiv:hep-ex/0502030}}.
[Erratum: \DOI{10.1103/PhysRevLett.100.049902}].

\bibitem{cdfxs}
\hrefCMSnoop {}{{CDF} Collaboration, ``{$\Upsilon$} Production and Polarization
  in p$\bar{\rm{p}}$ Collisions at $\sqrt{s}=$ 1.8 {TeV}'',} \textit{ Phys.
  Rev. Lett.} \textbf{ 88} (2002) 161802,
\href{http://dx.doi.org/10.1103/PhysRevLett.88.161802}{\doi{10.1103/PhysRevLett.88.161802}}.

\bibitem{NRQCD1}
\hrefCMSnoop {}{P.~L. Cho and A.~K. Leibovich, ``{Color octet quarkonia
  production}'',} \textit{ Phys. Rev. D} \textbf{ 53} (1996) 150,
  \href{http://dx.doi.org/10.1103/PhysRevD.53.150}{\doi{10.1103/PhysRevD.53.150}},
\href{http://www.arXiv.org/abs/hep-ph/9505329}{\texttt{arXiv:hep-ph/9505329}}.

\bibitem{NRQCD2}
\hrefCMSnoop {}{P.~L. Cho and A.~K. Leibovich, ``Color octet quarkonia
  production. {II}'',} \textit{ Phys. Rev. D} \textbf{ 53} (1996) 6203,
  \href{http://dx.doi.org/10.1103/PhysRevD.53.6203}{\doi{10.1103/PhysRevD.53.6203}},
\href{http://www.arXiv.org/abs/hep-ph/9511315}{\texttt{arXiv:hep-ph/9511315}}.

\bibitem{wang}
\hrefCMSnoop {}{K.~Wang, Y.-Q. Ma, and K.-T. Chao, ``{$\Upsilon(1S)$ prompt
  production at the Tevatron and LHC in nonrelativistic QCD}'',} \textit{ Phys.
  Rev. D} \textbf{ 85} (2012) 114003,
  \href{http://dx.doi.org/10.1103/PhysRevD.85.114003}{\doi{10.1103/PhysRevD.85.114003}},
\href{http://www.arXiv.org/abs/1202.6012}{\texttt{arXiv:1202.6012}}.

\bibitem{whitepaper}
G.~T. Bodwin\hrefCMSnoop {}{ {et~al.}, ``Quarkonium at the Frontiers of High
  Energy Physics: A {Snowmass} White Paper'',} (2013).
\href{http://www.arXiv.org/abs/1307.7425}{\texttt{arXiv:1307.7425}}.

\bibitem{Bodwin}
\hrefCMSnoop {}{G.~T. Bodwin, H.~S. Chung, U.-R. Kim, and J.~Lee,
  ``Fragmentation Contributions to {J/$\psi$} Production at the {Tevatron} and
  the {LHC}'',} \textit{ Phys. Rev. Lett.} \textbf{ 113} (2014) 022001,
  \href{http://dx.doi.org/10.1103/PhysRevLett.113.022001}{\doi{10.1103/PhysRevLett.113.022001}},
\href{http://www.arXiv.org/abs/1403.3612}{\texttt{arXiv:1403.3612}}.

\bibitem{gong}
\hrefCMSnoop {}{B.~Gong, L.-P. Wan, J.-X. Wang, and H.-F. Zhang, ``Complete
  Next-to-Leading-Order Study on the Yield and Polarization of {$\Upsilon$(1S,
  2S, 3S)} at the {Tevatron} and {LHC}'',} \textit{ Phys. Rev. Lett.} \textbf{
  112} (2014) 032001,
  \href{http://dx.doi.org/10.1103/PhysRevLett.112.032001}{\doi{10.1103/PhysRevLett.112.032001}},
\href{http://www.arXiv.org/abs/1305.0748}{\texttt{arXiv:1305.0748}}.

\bibitem{CSM}
P.~Artoisenet\hrefCMSnoop {}{ {et~al.}, ``{$\Upsilon$} Production at {Fermilab
  Tevatron} and {LHC} Energies'',} \textit{ Phys. Rev. Lett.} \textbf{ 101}
  (2008) 152001,
  \href{http://dx.doi.org/10.1103/PhysRevLett.101.152001}{\doi{10.1103/PhysRevLett.101.152001}},
\href{http://www.arXiv.org/abs/0806.3282}{\texttt{arXiv:0806.3282}}.

\bibitem{baranov}
\hrefCMSnoop {}{S.~P. Baranov, ``{Prompt $\Upsilon(nS)$ production at the LHC
  in view of the $k_t$-factorization approach}'',} \textit{ Phys. Rev. D}
  \textbf{ 86} (2012) 054015,
\href{http://dx.doi.org/10.1103/PhysRevD.86.054015}{\doi{10.1103/PhysRevD.86.054015}}.

\bibitem{Faccioli:2014cqa}
P.~Faccioli\hrefCMSnoop {}{ {et~al.}, ``{Quarkonium production in the LHC era:
  a polarized perspective}'',} \textit{ Phys. Lett. B} \textbf{ 736} (2014) 98,
  \href{http://dx.doi.org/10.1016/j.physletb.2014.07.006}{\doi{10.1016/j.physletb.2014.07.006}},
\href{http://www.arXiv.org/abs/1403.3970}{\texttt{arXiv:1403.3970}}.

\bibitem{Chatrchyan:2012xi}
\hrefCMSnoop {}{{CMS} Collaboration, ``{Performance of CMS muon reconstruction
  in pp collision events at $\sqrt{s}=7$ TeV}'',} \textit{ JINST} \textbf{ 7}
  (2012) P10002,
  \href{http://dx.doi.org/10.1088/1748-0221/7/10/P10002}{\doi{10.1088/1748-0221/7/10/P10002}},
\href{http://www.arXiv.org/abs/1206.4071}{\texttt{arXiv:1206.4071}}.

\bibitem{cmsdet}
\hrefCMSnoop {}{{CMS} Collaboration, ``{The CMS experiment at the CERN LHC}'',}
  \textit{ JINST} \textbf{ 3} (2008) S08004,
\href{http://dx.doi.org/10.1088/1748-0221/3/08/S08004}{\doi{10.1088/1748-0221/3/08/S08004}}.

\bibitem{evtgen}
\hrefCMSnoop {}{D.~J. Lange, ``The {EvtGen} particle decay simulation
  package'',} \textit{ Nucl. Instrum. Meth. A} \textbf{ 462} (2001) 152,
\href{http://dx.doi.org/10.1016/S0168-9002(01)00089-4}{\doi{10.1016/S0168-9002(01)00089-4}}.

\bibitem{GEANT4}
\hrefCMSnoop {}{{GEANT4} Collaboration, ``{GEANT4}---a simulation toolkit'',}
  \textit{ Nucl. Instrum. Meth. A} \textbf{ 506} (2003) 250,
\href{http://dx.doi.org/10.1016/S0168-9002(03)01368-8}{\doi{10.1016/S0168-9002(03)01368-8}}.

\bibitem{tandp}
\hrefCMSnoop {}{{CMS} Collaboration, ``{Measurements of inclusive $W$ and $Z$
  cross sections in pp collisions at $\sqrt{s}=7$ TeV}'',} \textit{ JHEP}
  \textbf{ 01} (2011) 080,
  \href{http://dx.doi.org/10.1007/JHEP01(2011)080}{\doi{10.1007/JHEP01(2011)080}},
\href{http://www.arXiv.org/abs/1012.2466}{\texttt{arXiv:1012.2466}}.

\bibitem{PHOTOS1}
\hrefCMSnoop {}{E.~Barberio, B.~van Eijk, and Z.~W{\c{a}}s, ``{PHOTOS: A
  universal Monte Carlo for QED radiative corrections in decays}'',} \textit{
  Comput. Phys. Commun.} \textbf{ 66} (1991) 115,
\href{http://dx.doi.org/10.1016/0010-4655(91)90012-A}{\doi{10.1016/0010-4655(91)90012-A}}.

\bibitem{PHOTOSv2}
\hrefCMSnoop {}{E.~Barberio and Z.~W{\c{a}}s, ``{PHOTOS: A universal Monte
  Carlo for QED radiative corrections: Version 2.0}'',} \textit{ Comput. Phys.
  Commun.} \textbf{ 79} (1994) 291,
\href{http://dx.doi.org/10.1016/0010-4655(94)90074-4}{\doi{10.1016/0010-4655(94)90074-4}}.

\bibitem{Verkerke:2003ir}
\hrefCMSnoop {}{W.~Verkerke and D.~P. Kirkby, ``{The RooFit toolkit for data
  modeling}'',} in \textit{ 2003 Computing in High Energy and Nuclear Physics
  (CHEP03)}, p.~186.
\newblock 2003.
\newblock
  \href{http://www.arXiv.org/abs/physics/0306116}{\texttt{arXiv:physics/0306116}}.
\newblock eConf C0303241 (2003) MOLT007.

\bibitem{aic}
\hrefCMSnoop {}{H.~Akaike, ``{A new look at the statistical model
  identification}'',} \textit{ IEEE Transactions on Automatic Control} \textbf{
  19} (1974) 716,
  \href{http://dx.doi.org/10.1109/TAC.1974.1100705}{\doi{10.1109/TAC.1974.1100705}}.

\bibitem{aic1}
\hrefCMSnoop {}{K.~Burnham and D.~Anderson, ``{Multimodel Inference:
  Understanding AIC and BIC in Model Selection}'',} \textit{ Sociological
  Methods Research} \textbf{ 33} (2004) 261,
  \href{http://dx.doi.org/10.1177/0049124104268644}{\doi{10.1177/0049124104268644}}.

\bibitem{lumi}
\href {http://cds.cern.ch/record/1434360}{{CMS} Collaboration, ``Absolute
  Calibration of the Luminosity Measurement at {CMS}: Winter 2012 Update'',}
  CMS Physics Analysis Summary CMS-PAS-SMP-12-008, 2012.

\bibitem{Lafferty}
\hrefCMSnoop {}{G.~D. Lafferty and T.~R. Wyatt, ``{Where to stick your data
  points: The treatment of measurements within wide bins}'',} \textit{ Nucl.
  Instrum. Meth. A} \textbf{ 355} (1995) 541,
\href{http://dx.doi.org/10.1016/0168-9002(94)01112-5}{\doi{10.1016/0168-9002(94)01112-5}}.

\bibitem{cem}
\hrefCMSnoop {}{J.~F. Amundson, O.~J.~P. Eboli, E.~M. Gregores, and F.~Halzen,
  ``{Colorless states in perturbative QCD: Charmonium and rapidity gaps}'',}
  \textit{ Phys. Lett. B} \textbf{ 372} (1996) 127,
  \href{http://dx.doi.org/10.1016/0370-2693(96)00035-4}{\doi{10.1016/0370-2693(96)00035-4}},
\href{http://www.arXiv.org/abs/hep-ph/9512248}{\texttt{arXiv:hep-ph/9512248}}.

\bibitem{buten1}
\hrefCMSnoop {}{M.~Butenschoen and B.~A. Kniehl, ``{World data of $J/\psi$
  production consolidate NRQCD factorization at NLO}'',} \textit{ Phys. Rev. D}
  \textbf{ 84} (2011) 051501,
  \href{http://dx.doi.org/10.1103/PhysRevD.84.051501}{\doi{10.1103/PhysRevD.84.051501}},
\href{http://www.arXiv.org/abs/1105.0820}{\texttt{arXiv:1105.0820}}.

\bibitem{buten2}
\hrefCMSnoop {}{M.~Butenschoen and B.~A. Kniehl, ``{Reconciling $J/\psi$
  production at HERA, RHIC, Tevatron, and LHC with NRQCD factorization at
  next-to-leading order}'',} \textit{ Phys. Rev. Lett.} \textbf{ 106} (2011)
  022003,
  \href{http://dx.doi.org/10.1103/PhysRevLett.106.022003}{\doi{10.1103/PhysRevLett.106.022003}},
\href{http://www.arXiv.org/abs/1009.5662}{\texttt{arXiv:1009.5662}}.

\end{thebibliography}\endgroup
\ifthenelse{\boolean{cms@external}}{}{
    \clearpage
    \appendix
    \numberwithin{table}{section}
    \section{Supplementary material \label{app:tables}}
    \begin{table*}[htp]
\centering
\caption{ The $\pt$ bin width, the weighted mean $\pt$ within a bin, and the differential cross section times the dimuon branching fraction: $(\rd\sigma/\rd\pt) \mathcal{B}(\PgUn\to \Pgmp\Pgmm )$ for the $\PgUa$, $\PgUb$, and $\PgUc$ with $0 < \abs{y} < 0.6$.  The statistical uncertainty in the differential cross section is given as the percentage of the cross section in the format: $\sigma_{\text{stat}}/(\rd\sigma/\rd\pt)\,(\%)$ and similarly for the systematic uncertainty.  The percentage systematic uncertainty for a negative systematic shift is given in parentheses.  The statistical uncertainties are derived from the fit to the dimuon mass spectrum.  The systematic uncertainties are discussed in the text.  The 2.2\% systematic uncertainty in the integrated luminosity is not included.  \label{xs_table_y0}}

\small
\setlength{\extrarowheight}{5pt}
\setlength{\tabcolsep}{4pt}
\begin{tabular}{cc|ccc|ccc|ccc}\hline
& &  \multicolumn{3}{c}{$\PgUa$} & \multicolumn{3}{c}{$\PgUb$} & \multicolumn{3}{c}{$\PgUc$} \\ \hline
\pt & $\langle \pt \rangle$ & $\frac{\rd\sigma}{\rd\pt} \mathcal{B}$  & $\frac{\sigma_\text{stat}}{\rd\sigma/\rd\pt}$   & $\frac{\sigma_\text{syst}}{\rd\sigma/\rd\pt}$ & $\frac{\rd\sigma}{\rd\pt} \mathcal{B}$  & $\frac{\sigma_\text{stat}}{\rd\sigma/\rd\pt}$   & $\frac{\sigma_\text{syst}}{\rd\sigma/\rd\pt}$ & $\frac{\rd\sigma}{\rd\pt} \mathcal{B}$  & $\frac{\sigma_\text{stat}}{\rd\sigma/\rd\pt}$   & $\frac{\sigma_\text{syst}}{\rd\sigma/\rd\pt}$    \\
 \GeVns{} & \GeVns{} & (fb/\GeVns) & (\%) & (\%) & (fb/\GeVns) & (\%) & (\%) & (fb/\GeVns) & (\%) & (\%) \\ \hline
10--12 & 11.0 & 60936 & 0.5 & 6.8 (7.2) & 20036 & 1.0 & 8.6 (9.4) & 10951 & 1.4 & 10.0 (11.3)\\
12--14 & 13.0 & 33828 & 0.5 & 5.1 (5.6) & 12072 & 1.1 & 6.0 (6.7) & 7181 & 1.4 & 7.1 (8.0)\\
14--16 & 15.0 & 19670 & 0.6 & 4.8 (5.2) & 7621 & 1.2 & 5.2 (5.7) & 4596 & 1.6 & 5.7 (6.3)\\
16--18 & 17.0 & 11504 & 0.8 & 4.6 (4.9) & 4858 & 1.3 & 5.1 (5.5) & 3097 & 1.7 & 5.6 (6.1)\\
18--20 & 19.0 & 6914 & 0.9 & 4.5 (4.7) & 3034 & 1.5 & 4.9 (5.2) & 1966 & 2.0 & 5.2 (5.6)\\
20--22 & 21.0 & 4235 & 1.1 & 4.2 (4.4) & 1932 & 1.8 & 4.6 (4.9) & 1423 & 2.2 & 5.0 (5.2)\\
22--24 & 23.0 & 2721 & 1.4 & 4.1 (4.3) & 1203 & 2.3 & 4.5 (4.6) & 893 & 2.7 & 4.7 (4.9)\\
24--26 & 25.0 & 1848 & 1.7 & 4.1 (4.2) & 849 & 2.6 & 4.6 (4.7) & 594 & 3.3 & 4.9 (5.1)\\
26--28 & 27.0 & 1117 & 2.2 & 4.3 (4.4) & 537 & 3.6 & 5.7 (5.8) & 381 & 4.4 & 6.1 (6.2)\\
28--30 & 29.0 & 845 & 2.5 & 4.4 (4.5) & 413 & 4.0 & 5.8 (5.8) & 314 & 4.7 & 6.0 (6.1)\\
30--32 & 31.0 & 593 & 2.8 & 4.3 (4.4) & 288 & 4.4 & 5.0 (5.1) & 209 & 5.2 & 5.0 (5.2)\\
32--34 & 33.0 & 420 & 3.5 & 4.7 (4.8) & 194 & 6.0 & 6.7 (6.8) & 162 & 6.5 & 6.4 (6.5)\\
34--36 & 35.0 & 314 & 3.8 & 4.6 (4.6) & 158 & 5.8 & 5.9 (6.0) & 112 & 7.1 & 5.7 (5.8)\\
36--38 & 37.0 & 209 & 4.8 & 4.5 (4.6) & 123 & 6.5 & 4.6 (4.7) & 92 & 7.5 & 4.6 (4.7)\\
38--40 & 39.0 & 157 & 5.4 & 4.2 (4.2) & 86 & 7.8 & 4.4 (4.6) & 61 & 9.5 & 4.6 (4.7)\\
40--43 & 41.4 & 114 & 5.0 & 4.0 (4.1) & 61 & 7.6 & 4.3 (4.4) & 42 & 9.3 & 4.4 (4.6)\\
43--46 & 44.4 & 76 & 6.3 & 5.8 (5.8) & 36 & 10.0 & 9.5 (9.5) & 39 & 9.4 & 7.8 (7.9)\\
46--50 & 47.9 & 49 & 6.6 & 4.2 (4.2) & 27 & 9.6 & 4.8 (4.9) & 22 & 10.9 & 5.1 (5.3)\\
50--55 & 52.4 & 24 & 9.2 & 5.9 (5.8) & 17 & 10.9 & 4.8 (4.9) & 12 & 13.6 & 4.4 (4.6)\\
55--60 & 57.4 & 15 & 11.0 & 6.3 (6.2) & 9.8 & 14.8 & 5.0 (5.1) & 8.5 & 15.5 & 4.6 (4.8)\\
60--70 & 64.6 & 9.2 & 10.1 & 6.6 (6.5) & 4.7 & 15.7 & 7.3 (7.3) & 4.9 & 14.8 & 6.0 (6.1)\\
70--100 & 82.0 & 2.3 & 12.6 & 11.9 (11.8) & 1.2 & 17.6 & 8.9 (8.9) & 1.0 & 19.1 & 7.3 (7.2)\\
\hline
\end{tabular}
\end{table*}

\begin{table*}[ht]
\centering
\topcaption{ The $\pt$ bin width, the weighted mean $\pt$ within a bin, and the differential cross section times the dimuon branching fraction: $(\rd\sigma/\pt) \mathcal{B}(\PgUn\to \Pgmp\Pgmm )$ for the $\PgUa$, $\PgUb$, and $\PgUc$ with $0.6 < \abs{y} < 1.2$.  The notation is the same as for Table~\ref{xs_table_y0}. \label{xs_table_y1}}

\small
\setlength{\extrarowheight}{5pt}
\setlength{\tabcolsep}{4pt}
\begin{tabular}{cc|ccc|ccc|ccc}\hline
& &  \multicolumn{3}{c}{$\PgUa$} & \multicolumn{3}{c}{$\PgUb$} & \multicolumn{3}{c}{$\PgUc$} \\ \hline
\pt & $\langle \pt \rangle$ & $\frac{\rd\sigma}{\rd\pt} \mathcal{B}$  & $\frac{\sigma_\text{stat}}{\rd\sigma/\rd\pt}$   & $\frac{\sigma_\text{syst}}{\rd\sigma/\rd\pt}$ & $\frac{\rd\sigma}{\rd\pt} \mathcal{B}$  & $\frac{\sigma_\text{stat}}{\rd\sigma/\rd\pt}$   & $\frac{\sigma_\text{syst}}{\rd\sigma/\rd\pt}$ & $\frac{\rd\sigma}{\rd\pt} \mathcal{B}$  & $\frac{\sigma_\text{stat}}{\rd\sigma/\rd\pt}$   & $\frac{\sigma_\text{syst}}{\rd\sigma/\rd\pt}$    \\
 \GeVns{} & \GeVns{} & (fb/\GeVns) & (\%) & (\%) & (fb/\GeVns) & (\%) & (\%) & (fb/\GeVns) & (\%) & (\%) \\ \hline
10--12 & 11.0 & 55260 & 0.5 & 6.1 (6.3) & 18371 & 1.2 & 7.8 (8.2) & 9855 & 1.6 & 8.9 (10.0)\\
12--14 & 13.0 & 31331 & 0.6 & 4.1 (4.2) & 10973 & 1.1 & 4.9 (5.2) & 6741 & 1.5 & 5.5 (6.0)\\
14--16 & 15.0 & 18063 & 0.7 & 3.8 (4.0) & 6685 & 1.5 & 4.6 (4.8) & 4298 & 2.0 & 5.5 (5.7)\\
16--18 & 17.0 & 10481 & 0.9 & 3.8 (3.9) & 4105 & 1.7 & 4.9 (5.1) & 2759 & 2.3 & 4.8 (5.0)\\
18--20 & 19.0 & 6286 & 1.1 & 3.7 (3.8) & 2624 & 2.1 & 4.6 (4.8) & 1871 & 2.6 & 4.6 (4.7)\\
20--22 & 21.0 & 3875 & 1.4 & 3.8 (3.9) & 1746 & 2.4 & 4.7 (4.8) & 1220 & 3.0 & 5.0 (5.2)\\
22--24 & 23.0 & 2439 & 1.6 & 3.6 (3.7) & 1188 & 2.4 & 3.8 (3.9) & 819 & 3.0 & 4.0 (4.2)\\
24--26 & 25.0 & 1633 & 1.9 & 3.8 (3.9) & 865 & 2.8 & 4.0 (4.1) & 530 & 3.7 & 4.0 (4.2)\\
26--28 & 27.0 & 1080 & 2.2 & 3.8 (3.9) & 551 & 3.4 & 4.5 (4.5) & 377 & 4.3 & 4.7 (4.8)\\
28--30 & 29.0 & 765 & 2.5 & 3.7 (3.8) & 397 & 3.9 & 3.9 (4.0) & 274 & 5.0 & 4.2 (4.3)\\
30--32 & 31.0 & 484 & 3.8 & 5.0 (5.1) & 229 & 6.5 & 9.0 (9.0) & 170 & 7.8 & 10.3 (10.3)\\
32--34 & 33.0 & 371 & 3.7 & 6.0 (6.1) & 175 & 6.0 & 9.8 (9.8) & 123 & 7.8 & 10.3 (10.4)\\
34--36 & 35.0 & 263 & 4.3 & 3.7 (3.8) & 151 & 6.1 & 4.0 (4.1) & 101 & 8.1 & 4.3 (4.4)\\
36--38 & 37.0 & 193 & 4.9 & 3.6 (3.7) & 113 & 7.0 & 3.9 (4.0) & 84 & 8.6 & 4.3 (4.4)\\
38--40 & 39.0 & 152 & 5.7 & 3.7 (3.8) & 87 & 8.1 & 4.0 (4.1) & 47 & 11.9 & 4.2 (4.3)\\
40--43 & 41.4 & 98 & 6.0 & 5.0 (5.0) & 55 & 8.4 & 4.3 (4.4) & 43 & 9.8 & 4.2 (4.3)\\
43--46 & 44.4 & 73 & 6.5 & 7.1 (7.1) & 37 & 10.3 & 9.9 (9.9) & 34 & 11.1 & 13.4 (13.4)\\
46--50 & 47.9 & 51 & 6.7 & 4.0 (4.0) & 23 & 11.0 & 3.9 (4.0) & 16 & 14.2 & 4.0 (4.2)\\
50--55 & 52.4 & 27 & 8.6 & 5.1 (5.1) & 14 & 13.2 & 4.5 (4.5) & 8.7 & 18.3 & 4.0 (4.2)\\
55--60 & 57.4 & 15 & 11.2 & 5.1 (5.0) & 9.2 & 15.5 & 4.6 (4.6) & 11 & 13.7 & 4.5 (4.7)\\
60--70 & 64.6 & 8.1 & 10.8 & 5.6 (5.6) & 3.6 & 19.7 & 5.1 (5.1) & 3.3 & 20.9 & 5.2 (5.3)\\
70--100 & 82.0 & 1.6 & 15.0 & 7.4 (7.2) & 1.0 & 24.0 & 7.0 (6.9) & 1.3 & 19.8 & 6.6 (6.7)\\
\hline
\end{tabular}
\end{table*}

\begin{table*}[ht]
\centering
\topcaption{ The $\pt$ bin width, the weighted mean $\pt$ within a bin, and the differential cross section times the dimuon branching fraction: $(\rd\sigma/\pt) \mathcal{B}(\PgUn\to \Pgmp\Pgmm )$ for the $\PgUa$, $\PgUb$, and $\PgUc$ with 0 $< \abs{y} < 1.2$.  The notation is the same as for Table~\ref{xs_table_y0}. \label{xs_table_y2}}

\small
\setlength{\extrarowheight}{5pt}
\setlength{\tabcolsep}{4pt}
\begin{tabular}{cc|ccc|ccc|ccc}\hline
& &  \multicolumn{3}{c}{$\PgUa$} & \multicolumn{3}{c}{$\PgUb$} & \multicolumn{3}{c}{$\PgUc$} \\ \hline
\pt & $\langle \pt \rangle$ & $\frac{\rd\sigma}{\rd\pt} \mathcal{B}$  & $\frac{\sigma_\text{stat}}{\rd\sigma/\rd\pt}$   & $\frac{\sigma_\text{syst}}{\rd\sigma/\rd\pt}$ & $\frac{\rd\sigma}{\rd\pt} \mathcal{B}$  & $\frac{\sigma_\text{stat}}{\rd\sigma/\rd\pt}$   & $\frac{\sigma_\text{syst}}{\rd\sigma/\rd\pt}$ & $\frac{\rd\sigma}{\rd\pt} \mathcal{B}$  & $\frac{\sigma_\text{stat}}{\rd\sigma/\rd\pt}$   & $\frac{\sigma_\text{syst}}{\rd\sigma/\rd\pt}$    \\
 \GeVns{} & \GeVns{} & (fb/\GeVns) & (\%) & (\%) & (fb/\GeVns) & (\%) & (\%) & (fb/\GeVns) & (\%) & (\%) \\ \hline
10--12 & 11.0 & 116415 & 0.4 & 6.4 (6.7) & 38540 & 0.7 & 8.1 (8.7) & 20882 & 1.1 & 9.3 (10.5)\\
12--14 & 13.0 & 65266 & 0.4 & 4.6 (4.9) & 23088 & 0.8 & 5.4 (5.9) & 13947 & 1.0 & 6.2 (6.9)\\
14--16 & 15.0 & 37778 & 0.5 & 4.3 (4.6) & 14321 & 0.9 & 4.7 (5.1) & 8909 & 1.3 & 5.2 (5.7)\\
16--18 & 17.0 & 22008 & 0.6 & 4.1 (4.4) & 8969 & 1.1 & 4.7 (5.0) & 5873 & 1.4 & 5.0 (5.3)\\
18--20 & 19.0 & 13212 & 0.7 & 4.1 (4.2) & 5665 & 1.3 & 4.6 (4.8) & 3842 & 1.6 & 4.7 (5.0)\\
20--22 & 21.0 & 8116 & 0.9 & 4.0 (4.1) & 3683 & 1.5 & 4.4 (4.6) & 2648 & 1.8 & 4.7 (4.9)\\
22--24 & 23.0 & 5162 & 1.0 & 3.9 (4.0) & 2393 & 1.7 & 4.1 (4.2) & 1713 & 2.0 & 4.3 (4.5)\\
24--26 & 25.0 & 3483 & 1.3 & 3.9 (4.0) & 1715 & 1.9 & 4.2 (4.3) & 1124 & 2.4 & 4.4 (4.5)\\
26--28 & 27.0 & 2197 & 1.5 & 4.0 (4.0) & 1089 & 2.5 & 4.6 (4.7) & 759 & 3.1 & 4.9 (5.0)\\
28--30 & 29.0 & 1611 & 1.8 & 4.0 (4.1) & 811 & 2.8 & 4.6 (4.7) & 588 & 3.4 & 4.8 (5.0)\\
30--32 & 31.0 & 1077 & 2.3 & 4.2 (4.3) & 517 & 3.8 & 5.7 (5.7) & 380 & 4.5 & 6.2 (6.3)\\
32--34 & 33.0 & 791 & 2.5 & 4.7 (4.7) & 369 & 4.2 & 6.5 (6.6) & 286 & 5.0 & 6.5 (6.6)\\
34--36 & 35.0 & 577 & 2.8 & 4.1 (4.1) & 309 & 4.2 & 4.7 (4.7) & 213 & 5.3 & 4.8 (4.9)\\
36--38 & 37.0 & 402 & 3.4 & 4.0 (4.1) & 236 & 4.8 & 4.2 (4.3) & 176 & 5.7 & 4.4 (4.6)\\
38--40 & 39.0 & 308 & 3.9 & 3.9 (3.9) & 173 & 5.6 & 4.2 (4.3) & 109 & 7.5 & 4.4 (4.5)\\
40--43 & 41.4 & 212 & 3.9 & 4.1 (4.2) & 116 & 5.6 & 4.2 (4.3) & 86 & 6.8 & 4.3 (4.5)\\
43--46 & 44.4 & 148 & 4.5 & 5.3 (5.3) & 73 & 7.2 & 7.5 (7.5) & 73 & 7.2 & 8.1 (8.2)\\
46--50 & 47.9 & 100 & 4.7 & 4.0 (4.0) & 50 & 7.3 & 4.2 (4.3) & 38 & 8.7 & 4.5 (4.6)\\
50--55 & 52.4 & 51 & 6.3 & 5.0 (5.0) & 31 & 8.4 & 4.5 (4.6) & 21 & 11.0 & 4.2 (4.4)\\
55--60 & 57.4 & 29 & 7.9 & 5.4 (5.4) & 19 & 10.7 & 4.8 (4.8) & 20 & 10.3 & 4.6 (4.7)\\
60--70 & 64.6 & 17 & 7.4 & 6.0 (5.9) & 8.3 & 12.4 & 6.1 (6.1) & 8.1 & 12.2 & 5.5 (5.6)\\
70--100 & 82.0 & 3.9 & 9.7 & 8.9 (8.8) & 2.2 & 14.6 & 7.6 (7.5) & 2.3 & 14.0 & 6.9 (6.9)\\
\hline
\end{tabular}
\end{table*}

\begin{table}[ht]
\centering
\topcaption{The \pt bin width and corrected yield ratios $R_{21}$ and $R_{31}$ for $\abs{y} <0.6$. The statistical uncertainty in the corrected yield ratios is given as the percentage of the ratio in the format: $\sigma_{\text{stat}}/R_{n1}\, (\%)$ ($n$ = 2,3) and similarly for the systematic uncertainty. The statistical uncertainties are derived from the fit to the dimuon mass spectrum. The systematic uncertainties are discussed in the text.  The 2.2\% systematic uncertainty in the integrated luminosity is not included. \label{RatioTabley0}}

\small
\setlength{\extrarowheight}{5pt}
\setlength{\tabcolsep}{4pt}
\begin{tabular}{c|ccc|ccc}\hline
\pt & $R_{21}$ & $\frac{\sigma_\text{stat}}{R_{21}}$   & $\frac{\sigma_\text{syst}}{R_{21}}$ & $R_{31}$  & $\frac{\sigma_\text{stat}}{R_{31}}$   & $\frac{\sigma_\text{syst}}{R_{31}}$ \\
\GeVns &  & (\%) & (\%) &  & (\%) & (\%)  \\ \hline
10--12 & 0.33 & 1.0 & 7.2  & 0.18 & 1.4 & 8.6 \\
12--14 & 0.36 & 1.1 & 5.9  & 0.21 & 1.5 & 6.7 \\
14--16 & 0.39 & 1.3 & 4.9  & 0.23 & 1.7 & 5.5 \\
16--18 & 0.42 & 1.4 & 4.7  & 0.27 & 1.8 & 5.2 \\
18--20 & 0.44 & 1.7 & 4.1  & 0.28 & 2.1 & 4.6 \\
20--22 & 0.46 & 2.1 & 3.8  & 0.34 & 2.4 & 4.2 \\
22--24 & 0.44 & 2.6 & 3.4  & 0.33 & 3.0 & 3.8 \\
24--26 & 0.46 & 3.0 & 3.4  & 0.32 & 3.6 & 3.8 \\
26--28 & 0.48 & 3.9 & 3.9  & 0.34 & 4.6 & 4.5 \\
28--30 & 0.49 & 4.4 & 3.8  & 0.37 & 5.0 & 4.1 \\
30--32 & 0.49 & 5.1 & 3.5  & 0.35 & 5.8 & 3.6 \\
32--34 & 0.46 & 6.4 & 4.1  & 0.39 & 6.9 & 4.1 \\
34--36 & 0.50 & 6.8 & 3.6  & 0.36 & 7.8 & 3.7 \\
36--38 & 0.59 & 7.8 & 3.3  & 0.44 & 8.6 & 4.0 \\
38--40 & 0.55 & 9.2 & 3.1  & 0.39 & 10.6 & 3.5 \\
40--43 & 0.54 & 8.8 & 3.0  & 0.37 & 10.3 & 3.3 \\
43--46 & 0.48 & 11.4 & 5.3  & 0.51 & 10.9 & 3.9 \\
46--50 & 0.55 & 11.3 & 3.1  & 0.45 & 12.3 & 3.5 \\
50--55 & 0.72 & 13.4 & 3.2  & 0.51 & 15.6 & 4.6 \\
55--60 & 0.66 & 17.4 & 2.9  & 0.57 & 18.2 & 4.8 \\
60--70 & 0.51 & 17.8 & 2.5  & 0.53 & 16.9 & 2.8 \\
70--100 & 0.53 & 20.3 & 4.1  & 0.44 & 21.9 & 9.1 \\
\hline
\end{tabular}
\end{table}

\begin{table}[ht]
\topcaption{The \pt bin width and corrected yield ratios $R_{21}$ and $R_{31}$ for 0.6 $< \abs{y} < 1.2$. The notation is the same as for Table~\ref{RatioTabley0}. \label{RatioTabley1}}
\centering

\small
\setlength{\extrarowheight}{5pt}
\setlength{\tabcolsep}{4pt}
\begin{tabular}{c|ccc|ccc}\hline
\pt & $R_{21}$ & $\frac{\sigma_\text{stat}}{R_{21}}$   & $\frac{\sigma_\text{syst}}{R_{21}}$ & $R_{31}$  & $\frac{\sigma_\text{stat}}{R_{31}}$   & $\frac{\sigma_\text{syst}}{R_{31}}$ \\
\GeVns &  & (\%) & (\%) &  & (\%) & (\%)  \\ \hline
10--12 & 0.33 & 1.1 & 5.5  & 0.18 & 1.6 & 6.4 \\
12--14 & 0.35 & 1.2 & 3.9  & 0.22 & 1.5 & 4.3 \\
14--16 & 0.37 & 1.4 & 3.6  & 0.24 & 1.9 & 4.1 \\
16--18 & 0.39 & 1.7 & 3.4  & 0.26 & 2.2 & 4.1 \\
18--20 & 0.42 & 2.1 & 3.4  & 0.30 & 2.5 & 3.6 \\
20--22 & 0.45 & 2.4 & 3.2  & 0.31 & 3.0 & 3.8 \\
22--24 & 0.49 & 2.8 & 2.8  & 0.34 & 3.3 & 3.0 \\
24--26 & 0.53 & 3.2 & 2.8  & 0.32 & 4.0 & 3.3 \\
26--28 & 0.51 & 3.9 & 3.1  & 0.35 & 4.7 & 3.2 \\
28--30 & 0.52 & 4.6 & 2.9  & 0.36 & 5.5 & 3.4 \\
30--32 & 0.47 & 6.4 & 5.3  & 0.35 & 7.6 & 6.6 \\
32--34 & 0.47 & 6.8 & 5.3  & 0.33 & 8.2 & 5.9 \\
34--36 & 0.58 & 7.3 & 3.0  & 0.39 & 8.9 & 3.5 \\
36--38 & 0.58 & 8.2 & 2.9  & 0.44 & 9.5 & 3.4 \\
38--40 & 0.57 & 9.4 & 2.9  & 0.31 & 12.8 & 3.2 \\
40--43 & 0.56 & 9.8 & 3.0  & 0.44 & 11.0 & 4.5 \\
43--46 & 0.51 & 12.9 & 6.8  & 0.47 & 15.5 & 5.4 \\
46--50 & 0.45 & 12.3 & 2.6  & 0.31 & 15.0 & 3.3 \\
50--55 & 0.51 & 14.9 & 2.6  & 0.33 & 19.4 & 4.2 \\
55--60 & 0.63 & 18.4 & 4.6  & 0.77 & 16.6 & 2.9 \\
60--70 & 0.46 & 21.0 & 4.7  & 0.38 & 23.2 & 6.4 \\
70--100 & 0.66 & 25.8 & 4.2  & 0.75 & 23.4 & 9.7 \\
\hline
\end{tabular}
\end{table}

\begin{table}[ht]
\centering

\small
\setlength{\extrarowheight}{5pt}
\setlength{\tabcolsep}{4pt}
\topcaption{The \pt bin width and corrected yield ratios $R_{21}$ and $R_{31}$ for $\abs{y} <1.2$. The notation is the same as for Table~\ref{RatioTabley0}. \label{RatioTabley2} }
\begin{tabular}{c|ccc|ccc}\hline
\pt & $R_{21}$ & $\frac{\sigma_\text{stat}}{R_{21}}$   & $\frac{\sigma_\text{syst}}{R_{21}}$ & $R_{31}$  & $\frac{\sigma_\text{stat}}{R_{31}}$   & $\frac{\sigma_\text{syst}}{R_{31}}$ \\
\GeVns &  & (\%) & (\%) &  & (\%) & (\%)  \\ \hline
10--12 & 0.33 & 0.7 & 6.2  & 0.18 & 1.0 & 7.4 \\
12--14 & 0.35 & 0.8 & 4.8  & 0.21 & 1.1 & 5.4 \\
14--16 & 0.38 & 1.0 & 4.1  & 0.24 & 1.3 & 4.5 \\
16--18 & 0.41 & 1.1 & 4.0  & 0.27 & 1.4 & 4.3 \\
18--20 & 0.43 & 1.3 & 3.6  & 0.29 & 1.6 & 3.9 \\
20--22 & 0.45 & 1.6 & 3.4  & 0.33 & 1.9 & 3.7 \\
22--24 & 0.46 & 1.9 & 3.0  & 0.33 & 2.2 & 3.3 \\
24--26 & 0.49 & 2.2 & 3.0  & 0.32 & 2.7 & 3.4 \\
26--28 & 0.50 & 2.8 & 3.2  & 0.34 & 3.3 & 3.4 \\
28--30 & 0.50 & 3.2 & 3.1  & 0.36 & 3.7 & 3.5 \\
30--32 & 0.48 & 4.0 & 3.3  & 0.35 & 4.6 & 3.5 \\
32--34 & 0.47 & 4.7 & 3.8  & 0.36 & 5.3 & 4.0 \\
34--36 & 0.53 & 5.0 & 3.1  & 0.37 & 5.9 & 3.5 \\
36--38 & 0.58 & 5.7 & 3.1  & 0.44 & 6.4 & 3.4 \\
38--40 & 0.56 & 6.6 & 3.0  & 0.35 & 8.2 & 3.3 \\
40--43 & 0.55 & 6.6 & 2.9  & 0.40 & 7.5 & 3.2 \\
43--46 & 0.49 & 8.6 & 4.5  & 0.50 & 8.9 & 3.7 \\
46--50 & 0.49 & 8.4 & 2.7  & 0.37 & 9.7 & 3.2 \\
50--55 & 0.59 & 10.1 & 2.6  & 0.40 & 12.5 & 3.6 \\
55--60 & 0.65 & 12.6 & 2.8  & 0.65 & 12.4 & 3.0 \\
60--70 & 0.49 & 13.6 & 2.5  & 0.45 & 13.9 & 2.6 \\
70--100 & 0.56 & 16.0 & 3.3  & 0.50 & 16.5 & 7.0 \\
\hline
\end{tabular}
\end{table}

\begin{table}[ht]
\centering
\topcaption{The dimuon acceptance $\mathcal{A}$  calculated using the CMS measured polarization and its positive and negative uncertainties $\mathcal{A}(\sigma^{\pm})$, and the values of the acceptance assuming no polarization $\mathcal{A}$(unpol), transverse polarization $\mathcal{A}(T)$, and longitudinal polarization $\mathcal{A}(L)$ for the $\PgUa$ state and $\abs{y} <0.6$. \label{AccSumy01S}}
\small
\setlength{\extrarowheight}{5pt}
\setlength{\tabcolsep}{4pt}
\begin{tabular}{ccccccccc}
\hline
\pt [\GeVns{}] & $\mathcal{A}$ & $\mathcal{A}(\sigma^{+})$ & $\mathcal{A}(\sigma^{-})$ & $\mathcal{A}(\text{unpol})$ & $\mathcal{A}(T)$ & $\mathcal{A}(L)$ \\ \hline
10--12 &  0.31 & 0.29 & 0.32 & 0.32 & 0.26 & 0.45 \\
12--14 &  0.38 & 0.36 & 0.40 & 0.39 & 0.32 & 0.54 \\
14--16 &  0.44 & 0.42 & 0.45 & 0.45 & 0.36 & 0.61 \\
16--18 &  0.49 & 0.48 & 0.51 & 0.50 & 0.41 & 0.67 \\
18--20 &  0.53 & 0.52 & 0.55 & 0.54 & 0.45 & 0.71 \\
20--22 &  0.57 & 0.56 & 0.58 & 0.57 & 0.48 & 0.74 \\
22--24 &  0.60 & 0.59 & 0.61 & 0.60 & 0.51 & 0.77 \\
24--26 &  0.62 & 0.61 & 0.64 & 0.62 & 0.54 & 0.79 \\
26--28 &  0.64 & 0.63 & 0.66 & 0.65 & 0.56 & 0.81 \\
28--30 &  0.66 & 0.64 & 0.67 & 0.66 & 0.58 & 0.83 \\
30--32 &  0.67 & 0.66 & 0.69 & 0.68 & 0.60 & 0.84 \\
32--34 &  0.69 & 0.67 & 0.70 & 0.69 & 0.62 & 0.85 \\
34--36 &  0.70 & 0.69 & 0.71 & 0.71 & 0.63 & 0.86 \\
36--38 &  0.71 & 0.70 & 0.72 & 0.72 & 0.64 & 0.87 \\
38--40 &  0.72 & 0.71 & 0.74 & 0.73 & 0.66 & 0.88 \\
40--43 &  0.73 & 0.72 & 0.75 & 0.74 & 0.67 & 0.89 \\
43--46 &  0.75 & 0.73 & 0.76 & 0.76 & 0.69 & 0.89 \\
46--50 &  0.76 & 0.75 & 0.78 & 0.77 & 0.71 & 0.90 \\
50--55 &  0.78 & 0.77 & 0.80 & 0.79 & 0.73 & 0.91 \\
55--60 &  0.80 & 0.78 & 0.81 & 0.80 & 0.75 & 0.92 \\
60--70 &  0.82 & 0.81 & 0.83 & 0.82 & 0.77 & 0.93 \\
70--100 &  0.87 & 0.86 & 0.88 & 0.87 & 0.83 & 0.96 \\
\hline
\end{tabular}
\end{table}

\begin{table}[ht]
\centering
\topcaption{The dimuon acceptance $\mathcal{A}$  calculated using the CMS measured polarization and its positive and negative uncertainties $\mathcal{A}(\sigma^{\pm})$, and the values of the acceptance assuming no polarization $\mathcal{A}$(unpol), transverse polarization $\mathcal{A}(T)$, and longitudinal polarization $\mathcal{A}(L)$ for the $\PgUa$ state and $ 0.6 < \abs{y} < 1.2$. \label{AccSumy11S}}
\small
\setlength{\extrarowheight}{5pt}
\setlength{\tabcolsep}{4pt}
\begin{tabular}{ccccccccc}
\hline
\pt [\GeVns{}] & $\mathcal{A}$ & $\mathcal{A}(\sigma^{+})$ & $\mathcal{A}(\sigma^{-})$ & $\mathcal{A}(\text{unpol})$ & $\mathcal{A}(T)$ & $\mathcal{A}(L)$ \\ \hline
10--12 &  0.29 & 0.28 & 0.30 & 0.29 & 0.24 & 0.38 \\
12--14 &  0.35 & 0.34 & 0.36 & 0.35 & 0.30 & 0.46 \\
14--16 &  0.42 & 0.41 & 0.43 & 0.42 & 0.36 & 0.54 \\
16--18 &  0.47 & 0.46 & 0.48 & 0.47 & 0.40 & 0.60 \\
18--20 &  0.52 & 0.51 & 0.53 & 0.51 & 0.44 & 0.66 \\
20--22 &  0.55 & 0.54 & 0.56 & 0.55 & 0.48 & 0.70 \\
22--24 &  0.59 & 0.58 & 0.60 & 0.59 & 0.51 & 0.74 \\
24--26 &  0.61 & 0.60 & 0.63 & 0.62 & 0.54 & 0.77 \\
26--28 &  0.64 & 0.63 & 0.65 & 0.64 & 0.57 & 0.79 \\
28--30 &  0.66 & 0.64 & 0.67 & 0.66 & 0.59 & 0.81 \\
30--32 &  0.67 & 0.66 & 0.69 & 0.68 & 0.61 & 0.83 \\
32--34 &  0.69 & 0.67 & 0.70 & 0.70 & 0.63 & 0.84 \\
34--36 &  0.70 & 0.69 & 0.71 & 0.71 & 0.64 & 0.86 \\
36--38 &  0.71 & 0.70 & 0.72 & 0.73 & 0.66 & 0.87 \\
38--40 &  0.72 & 0.71 & 0.74 & 0.74 & 0.67 & 0.88 \\
40--43 &  0.74 & 0.73 & 0.75 & 0.75 & 0.68 & 0.88 \\
43--46 &  0.75 & 0.74 & 0.77 & 0.77 & 0.70 & 0.89 \\
46--50 &  0.77 & 0.76 & 0.78 & 0.78 & 0.72 & 0.90 \\
50--55 &  0.79 & 0.78 & 0.80 & 0.80 & 0.74 & 0.91 \\
55--60 &  0.81 & 0.80 & 0.82 & 0.81 & 0.76 & 0.92 \\
60--70 &  0.83 & 0.82 & 0.84 & 0.83 & 0.78 & 0.93 \\
70--100 &  0.88 & 0.87 & 0.89 & 0.88 & 0.84 & 0.96 \\
\hline
\end{tabular}
\end{table}

\begin{table}[ht]
\centering
\topcaption{The dimuon acceptance $\mathcal{A}$  calculated using the CMS measured polarization and its positive and negative uncertainties $\mathcal{A}(\sigma^{\pm})$, and the values of the acceptance assuming no polarization $\mathcal{A}$(unpol), transverse polarization $\mathcal{A}(T)$, and longitudinal polarization $\mathcal{A}(L)$ for the $\PgUa$ state and $\abs{y} <1.2$. \label{AccSumy21S}}
\small
\setlength{\extrarowheight}{5pt}
\setlength{\tabcolsep}{4pt}
\begin{tabular}{ccccccccc}
\hline
\pt [\GeVns{}] & $\mathcal{A}$ & $\mathcal{A}(\sigma^{+})$ & $\mathcal{A}(\sigma^{-})$ & $\mathcal{A}(\text{unpol})$ & $\mathcal{A}(T)$ & $\mathcal{A}(L)$ \\ \hline
10--12 &  0.30 & 0.28 & 0.31 & 0.30 & 0.25 & 0.42 \\
12--14 &  0.37 & 0.35 & 0.38 & 0.37 & 0.31 & 0.50 \\
14--16 &  0.43 & 0.42 & 0.44 & 0.43 & 0.36 & 0.57 \\
16--18 &  0.48 & 0.47 & 0.49 & 0.48 & 0.41 & 0.64 \\
18--20 &  0.52 & 0.51 & 0.54 & 0.53 & 0.45 & 0.68 \\
20--22 &  0.56 & 0.55 & 0.57 & 0.56 & 0.48 & 0.72 \\
22--24 &  0.59 & 0.58 & 0.61 & 0.59 & 0.51 & 0.75 \\
24--26 &  0.62 & 0.61 & 0.63 & 0.62 & 0.54 & 0.78 \\
26--28 &  0.64 & 0.63 & 0.65 & 0.64 & 0.56 & 0.80 \\
28--30 &  0.66 & 0.64 & 0.67 & 0.66 & 0.58 & 0.82 \\
30--32 &  0.67 & 0.66 & 0.69 & 0.68 & 0.60 & 0.84 \\
32--34 &  0.69 & 0.67 & 0.70 & 0.70 & 0.62 & 0.85 \\
34--36 &  0.70 & 0.69 & 0.71 & 0.71 & 0.64 & 0.86 \\
36--38 &  0.71 & 0.70 & 0.72 & 0.72 & 0.65 & 0.87 \\
38--40 &  0.72 & 0.71 & 0.74 & 0.73 & 0.66 & 0.88 \\
40--43 &  0.74 & 0.72 & 0.75 & 0.75 & 0.68 & 0.88 \\
43--46 &  0.75 & 0.74 & 0.77 & 0.76 & 0.70 & 0.89 \\
46--50 &  0.77 & 0.76 & 0.78 & 0.78 & 0.71 & 0.90 \\
50--55 &  0.79 & 0.77 & 0.80 & 0.79 & 0.73 & 0.91 \\
55--60 &  0.80 & 0.79 & 0.81 & 0.81 & 0.75 & 0.92 \\
60--70 &  0.82 & 0.81 & 0.84 & 0.83 & 0.78 & 0.93 \\
70--100 &  0.87 & 0.86 & 0.89 & 0.88 & 0.83 & 0.96 \\
\hline
\end{tabular}
\end{table}

\begin{table}[ht]
\centering
\topcaption{The dimuon acceptance $\mathcal{A}$  calculated using the CMS measured polarization and its positive and negative uncertainties $\mathcal{A}(\sigma^{\pm})$, and the values of the acceptance assuming no polarization $\mathcal{A}$(unpol), transverse polarization $\mathcal{A}(T)$, and longitudinal polarization $\mathcal{A}(L)$ for the $\PgUb$ state and $\abs{y} <0.6$. \label{AccSumy02S}}
\small
\setlength{\extrarowheight}{5pt}
\setlength{\tabcolsep}{4pt}
\begin{tabular}{ccccccccc}
\hline
\pt [\GeVns{}] & $\mathcal{A}$ & $\mathcal{A}(\sigma^{+})$ & $\mathcal{A}(\sigma^{-})$ & $\mathcal{A}(\text{unpol})$ & $\mathcal{A}(T)$ & $\mathcal{A}(L)$ \\ \hline
10--12 &  0.31 & 0.30 & 0.33 & 0.34 & 0.27 & 0.48 \\
12--14 &  0.37 & 0.36 & 0.39 & 0.41 & 0.33 & 0.56 \\
14--16 &  0.43 & 0.41 & 0.45 & 0.47 & 0.38 & 0.63 \\
16--18 &  0.48 & 0.46 & 0.50 & 0.51 & 0.42 & 0.68 \\
18--20 &  0.52 & 0.51 & 0.54 & 0.55 & 0.46 & 0.73 \\
20--22 &  0.56 & 0.54 & 0.57 & 0.58 & 0.49 & 0.76 \\
22--24 &  0.59 & 0.57 & 0.61 & 0.61 & 0.52 & 0.79 \\
24--26 &  0.61 & 0.59 & 0.63 & 0.63 & 0.55 & 0.81 \\
26--28 &  0.64 & 0.62 & 0.65 & 0.66 & 0.58 & 0.83 \\
28--30 &  0.65 & 0.63 & 0.67 & 0.67 & 0.59 & 0.84 \\
30--32 &  0.66 & 0.64 & 0.68 & 0.69 & 0.61 & 0.85 \\
32--34 &  0.68 & 0.66 & 0.70 & 0.71 & 0.63 & 0.87 \\
34--36 &  0.69 & 0.67 & 0.71 & 0.72 & 0.64 & 0.88 \\
36--38 &  0.70 & 0.68 & 0.72 & 0.73 & 0.65 & 0.88 \\
38--40 &  0.71 & 0.69 & 0.73 & 0.74 & 0.66 & 0.89 \\
40--43 &  0.73 & 0.71 & 0.75 & 0.76 & 0.69 & 0.90 \\
43--46 &  0.74 & 0.72 & 0.76 & 0.77 & 0.70 & 0.91 \\
46--50 &  0.76 & 0.74 & 0.78 & 0.79 & 0.72 & 0.92 \\
50--55 &  0.77 & 0.76 & 0.79 & 0.80 & 0.74 & 0.92 \\
55--60 &  0.79 & 0.78 & 0.81 & 0.82 & 0.76 & 0.93 \\
60--70 &  0.82 & 0.80 & 0.83 & 0.84 & 0.78 & 0.95 \\
70--100 &  0.87 & 0.85 & 0.88 & 0.88 & 0.84 & 0.97 \\
\hline
\end{tabular}
\end{table}

\begin{table}[ht]
\centering
\topcaption{The dimuon acceptance $\mathcal{A}$  calculated using the CMS measured polarization and its positive and negative uncertainties $\mathcal{A}(\sigma^{\pm})$, and the values of the acceptance assuming no polarization $\mathcal{A}$(unpol), transverse polarization $\mathcal{A}(T)$, and longitudinal polarization $\mathcal{A}(L)$ for the $\PgUb$ state and $ 0.6 < \abs{y} < 1.2$. \label{AccSumy12S}}
\small
\setlength{\extrarowheight}{5pt}
\setlength{\tabcolsep}{4pt}
\begin{tabular}{ccccccccc}
\hline
\pt [\GeVns{}] & $\mathcal{A}$ & $\mathcal{A}(\sigma^{+})$ & $\mathcal{A}(\sigma^{-})$ & $\mathcal{A}(\text{unpol})$ & $\mathcal{A}(T)$ & $\mathcal{A}(L)$ \\ \hline
10--12 &  0.28 & 0.27 & 0.30 & 0.30 & 0.25 & 0.39 \\
12--14 &  0.35 & 0.34 & 0.37 & 0.36 & 0.31 & 0.47 \\
14--16 &  0.42 & 0.41 & 0.43 & 0.42 & 0.36 & 0.54 \\
16--18 &  0.47 & 0.46 & 0.49 & 0.47 & 0.41 & 0.60 \\
18--20 &  0.51 & 0.50 & 0.53 & 0.52 & 0.45 & 0.66 \\
20--22 &  0.54 & 0.53 & 0.56 & 0.56 & 0.49 & 0.70 \\
22--24 &  0.56 & 0.55 & 0.58 & 0.59 & 0.52 & 0.74 \\
24--26 &  0.58 & 0.57 & 0.60 & 0.62 & 0.54 & 0.77 \\
26--28 &  0.61 & 0.60 & 0.63 & 0.65 & 0.57 & 0.79 \\
28--30 &  0.64 & 0.62 & 0.66 & 0.67 & 0.60 & 0.82 \\
30--32 &  0.66 & 0.64 & 0.68 & 0.69 & 0.61 & 0.84 \\
32--34 &  0.68 & 0.66 & 0.69 & 0.70 & 0.63 & 0.85 \\
34--36 &  0.69 & 0.68 & 0.71 & 0.72 & 0.65 & 0.87 \\
36--38 &  0.72 & 0.70 & 0.73 & 0.74 & 0.67 & 0.88 \\
38--40 &  0.72 & 0.71 & 0.74 & 0.75 & 0.68 & 0.89 \\
40--43 &  0.74 & 0.72 & 0.76 & 0.76 & 0.69 & 0.89 \\
43--46 &  0.76 & 0.74 & 0.78 & 0.78 & 0.71 & 0.91 \\
46--50 &  0.77 & 0.75 & 0.79 & 0.79 & 0.73 & 0.92 \\
50--55 &  0.79 & 0.78 & 0.81 & 0.81 & 0.75 & 0.93 \\
55--60 &  0.81 & 0.80 & 0.83 & 0.83 & 0.77 & 0.94 \\
60--70 &  0.83 & 0.82 & 0.85 & 0.84 & 0.79 & 0.95 \\
70--100 &  0.88 & 0.87 & 0.90 & 0.89 & 0.85 & 0.97 \\
\hline
\end{tabular}
\end{table}

\begin{table}[ht]
\centering
\topcaption{The dimuon acceptance $\mathcal{A}$  calculated using the CMS measured polarization and its positive and negative uncertainties $\mathcal{A}(\sigma^{\pm})$, and the values of the acceptance assuming no polarization $\mathcal{A}$(unpol), transverse polarization $\mathcal{A}(T)$, and longitudinal polarization $\mathcal{A}(L)$ for the $\PgUb$ state and $\abs{y} <1.2$. \label{AccSumy22S}}
\small
\setlength{\extrarowheight}{5pt}
\setlength{\tabcolsep}{4pt}
\begin{tabular}{ccccccccc}
\hline
\pt [\GeVns{}] & $\mathcal{A}$ & $\mathcal{A}(\sigma^{+})$ & $\mathcal{A}(\sigma^{-})$ & $\mathcal{A}(\text{unpol})$ & $\mathcal{A}(T)$ & $\mathcal{A}(L)$ \\ \hline
10--12 &  0.30 & 0.28 & 0.32 & 0.32 & 0.26 & 0.44 \\
12--14 &  0.36 & 0.35 & 0.38 & 0.38 & 0.32 & 0.52 \\
14--16 &  0.42 & 0.41 & 0.44 & 0.44 & 0.37 & 0.59 \\
16--18 &  0.47 & 0.46 & 0.49 & 0.49 & 0.41 & 0.64 \\
18--20 &  0.52 & 0.50 & 0.53 & 0.53 & 0.46 & 0.69 \\
20--22 &  0.55 & 0.53 & 0.56 & 0.57 & 0.49 & 0.73 \\
22--24 &  0.58 & 0.56 & 0.59 & 0.60 & 0.52 & 0.76 \\
24--26 &  0.60 & 0.58 & 0.61 & 0.62 & 0.54 & 0.79 \\
26--28 &  0.62 & 0.61 & 0.64 & 0.65 & 0.57 & 0.81 \\
28--30 &  0.64 & 0.63 & 0.66 & 0.67 & 0.59 & 0.83 \\
30--32 &  0.66 & 0.64 & 0.68 & 0.69 & 0.61 & 0.85 \\
32--34 &  0.68 & 0.66 & 0.70 & 0.71 & 0.63 & 0.86 \\
34--36 &  0.69 & 0.67 & 0.71 & 0.72 & 0.65 & 0.87 \\
36--38 &  0.71 & 0.69 & 0.72 & 0.73 & 0.66 & 0.88 \\
38--40 &  0.71 & 0.70 & 0.73 & 0.74 & 0.67 & 0.89 \\
40--43 &  0.73 & 0.72 & 0.75 & 0.76 & 0.69 & 0.90 \\
43--46 &  0.75 & 0.73 & 0.77 & 0.77 & 0.71 & 0.91 \\
46--50 &  0.76 & 0.75 & 0.78 & 0.79 & 0.72 & 0.92 \\
50--55 &  0.78 & 0.77 & 0.80 & 0.80 & 0.74 & 0.93 \\
55--60 &  0.80 & 0.79 & 0.82 & 0.82 & 0.76 & 0.93 \\
60--70 &  0.82 & 0.81 & 0.84 & 0.84 & 0.79 & 0.95 \\
70--100 &  0.87 & 0.86 & 0.89 & 0.88 & 0.84 & 0.97 \\
\hline
\end{tabular}
\end{table}

\begin{table}[ht]
\centering
\topcaption{The dimuon acceptance $\mathcal{A}$  calculated using the CMS measured polarization and its positive and negative uncertainties $\mathcal{A}(\sigma^{\pm})$, and the values of the acceptance assuming no polarization $\mathcal{A}$(unpol), transverse polarization $\mathcal{A}(T)$, and longitudinal polarization $\mathcal{A}(L)$ for the $\PgUc$ state and $\abs{y} <0.6$. \label{AccSumy03S}}
\small
\setlength{\extrarowheight}{5pt}
\setlength{\tabcolsep}{4pt}
\begin{tabular}{ccccccccc}
\hline
\pt [\GeVns{}] & $\mathcal{A}$ & $\mathcal{A}(\sigma^{+})$ & $\mathcal{A}(\sigma^{-})$ & $\mathcal{A}(\text{unpol})$ & $\mathcal{A}(T)$ & $\mathcal{A}(L)$ \\ \hline
10--12 &  0.34 & 0.32 & 0.37 & 0.35 & 0.28 & 0.49 \\
12--14 &  0.39 & 0.37 & 0.42 & 0.42 & 0.34 & 0.58 \\
14--16 &  0.44 & 0.42 & 0.46 & 0.47 & 0.39 & 0.64 \\
16--18 &  0.49 & 0.47 & 0.51 & 0.52 & 0.43 & 0.69 \\
18--20 &  0.52 & 0.50 & 0.55 & 0.55 & 0.46 & 0.73 \\
20--22 &  0.55 & 0.54 & 0.58 & 0.59 & 0.50 & 0.77 \\
22--24 &  0.59 & 0.57 & 0.60 & 0.62 & 0.53 & 0.79 \\
24--26 &  0.61 & 0.59 & 0.63 & 0.64 & 0.55 & 0.82 \\
26--28 &  0.63 & 0.62 & 0.66 & 0.67 & 0.58 & 0.84 \\
28--30 &  0.65 & 0.63 & 0.67 & 0.68 & 0.60 & 0.85 \\
30--32 &  0.67 & 0.65 & 0.69 & 0.70 & 0.61 & 0.86 \\
32--34 &  0.69 & 0.67 & 0.71 & 0.71 & 0.63 & 0.87 \\
34--36 &  0.70 & 0.68 & 0.72 & 0.73 & 0.65 & 0.88 \\
36--38 &  0.72 & 0.70 & 0.74 & 0.74 & 0.67 & 0.89 \\
38--40 &  0.73 & 0.70 & 0.75 & 0.75 & 0.67 & 0.90 \\
40--43 &  0.74 & 0.72 & 0.76 & 0.76 & 0.69 & 0.91 \\
43--46 &  0.75 & 0.73 & 0.78 & 0.78 & 0.71 & 0.91 \\
46--50 &  0.77 & 0.75 & 0.79 & 0.79 & 0.72 & 0.92 \\
50--55 &  0.79 & 0.77 & 0.81 & 0.81 & 0.74 & 0.93 \\
55--60 &  0.81 & 0.79 & 0.83 & 0.82 & 0.77 & 0.94 \\
60--70 &  0.82 & 0.81 & 0.85 & 0.84 & 0.79 & 0.95 \\
70--100 &  0.87 & 0.85 & 0.89 & 0.88 & 0.84 & 0.98 \\
\hline
\end{tabular}
\end{table}

\begin{table}[ht]
\centering
\topcaption{The dimuon acceptance $\mathcal{A}$  calculated using the CMS measured polarization and its positive and negative uncertainties $\mathcal{A}(\sigma^{\pm})$, and the values of the acceptance assuming no polarization $\mathcal{A}$(unpol), transverse polarization $\mathcal{A}(T)$, and longitudinal polarization $\mathcal{A}(L)$ for the $\PgUc$ state and $ 0.6 < \abs{y} < 1.2$. \label{AccSumy13S}}
\small
\setlength{\extrarowheight}{5pt}
\setlength{\tabcolsep}{4pt}
\begin{tabular}{ccccccccc}
\hline
\pt [\GeVns{}] & $\mathcal{A}$ & $\mathcal{A}(\sigma^{+})$ & $\mathcal{A}(\sigma^{-})$ & $\mathcal{A}(\text{unpol})$ & $\mathcal{A}(T)$ & $\mathcal{A}(L)$ \\ \hline
10--12 &  0.31 & 0.29 & 0.33 & 0.30 & 0.25 & 0.40 \\
12--14 &  0.35 & 0.34 & 0.37 & 0.36 & 0.31 & 0.47 \\
14--16 &  0.40 & 0.39 & 0.41 & 0.42 & 0.36 & 0.54 \\
16--18 &  0.44 & 0.43 & 0.46 & 0.48 & 0.41 & 0.60 \\
18--20 &  0.49 & 0.48 & 0.50 & 0.52 & 0.46 & 0.66 \\
20--22 &  0.53 & 0.52 & 0.55 & 0.56 & 0.49 & 0.70 \\
22--24 &  0.58 & 0.56 & 0.59 & 0.59 & 0.52 & 0.73 \\
24--26 &  0.61 & 0.60 & 0.63 & 0.62 & 0.55 & 0.77 \\
26--28 &  0.63 & 0.61 & 0.65 & 0.64 & 0.57 & 0.79 \\
28--30 &  0.65 & 0.63 & 0.67 & 0.67 & 0.60 & 0.82 \\
30--32 &  0.66 & 0.65 & 0.68 & 0.69 & 0.62 & 0.84 \\
32--34 &  0.68 & 0.66 & 0.70 & 0.71 & 0.64 & 0.85 \\
34--36 &  0.69 & 0.67 & 0.71 & 0.72 & 0.65 & 0.87 \\
36--38 &  0.70 & 0.68 & 0.72 & 0.74 & 0.67 & 0.88 \\
38--40 &  0.72 & 0.70 & 0.74 & 0.76 & 0.69 & 0.89 \\
40--43 &  0.73 & 0.72 & 0.76 & 0.77 & 0.70 & 0.90 \\
43--46 &  0.75 & 0.73 & 0.77 & 0.78 & 0.72 & 0.91 \\
46--50 &  0.77 & 0.75 & 0.79 & 0.80 & 0.73 & 0.92 \\
50--55 &  0.79 & 0.77 & 0.81 & 0.81 & 0.75 & 0.93 \\
55--60 &  0.81 & 0.79 & 0.83 & 0.83 & 0.77 & 0.94 \\
60--70 &  0.83 & 0.82 & 0.85 & 0.85 & 0.80 & 0.95 \\
70--100 &  0.88 & 0.87 & 0.90 & 0.89 & 0.85 & 0.98 \\
\hline
\end{tabular}
\end{table}

\begin{table}[ht]
\centering
\topcaption{The dimuon acceptance $\mathcal{A}$  calculated using the CMS measured polarization and its positive and negative uncertainties $\mathcal{A}(\sigma^{\pm})$, and the values of the acceptance assuming no polarization $\mathcal{A}$(unpol), transverse polarization $\mathcal{A}(T)$, and longitudinal polarization $\mathcal{A}(L)$ for the $\PgUc$ state and $\abs{y} < 1.2$. \label{AccSumy23S}}
\small
\setlength{\extrarowheight}{5pt}
\setlength{\tabcolsep}{4pt}
\begin{tabular}{ccccccccc}
\hline
\pt [\GeVns{}] & $\mathcal{A}$ & $\mathcal{A}(\sigma^{+})$ & $\mathcal{A}(\sigma^{-})$ & $\mathcal{A}(\text{unpol})$ & $\mathcal{A}(T)$ & $\mathcal{A}(L)$ \\ \hline
10--12 &  0.32 & 0.30 & 0.35 & 0.33 & 0.27 & 0.44 \\
12--14 &  0.37 & 0.36 & 0.39 & 0.39 & 0.33 & 0.52 \\
14--16 &  0.42 & 0.40 & 0.44 & 0.45 & 0.37 & 0.59 \\
16--18 &  0.47 & 0.45 & 0.48 & 0.50 & 0.42 & 0.65 \\
18--20 &  0.51 & 0.49 & 0.52 & 0.54 & 0.46 & 0.69 \\
20--22 &  0.54 & 0.53 & 0.56 & 0.57 & 0.49 & 0.73 \\
22--24 &  0.58 & 0.57 & 0.60 & 0.60 & 0.53 & 0.76 \\
24--26 &  0.61 & 0.59 & 0.63 & 0.63 & 0.55 & 0.79 \\
26--28 &  0.63 & 0.61 & 0.65 & 0.65 & 0.58 & 0.81 \\
28--30 &  0.65 & 0.63 & 0.67 & 0.68 & 0.60 & 0.83 \\
30--32 &  0.67 & 0.65 & 0.69 & 0.69 & 0.62 & 0.85 \\
32--34 &  0.68 & 0.66 & 0.70 & 0.71 & 0.63 & 0.86 \\
34--36 &  0.69 & 0.67 & 0.72 & 0.72 & 0.65 & 0.87 \\
36--38 &  0.71 & 0.69 & 0.73 & 0.74 & 0.67 & 0.89 \\
38--40 &  0.72 & 0.70 & 0.75 & 0.75 & 0.68 & 0.89 \\
40--43 &  0.74 & 0.72 & 0.76 & 0.77 & 0.70 & 0.90 \\
43--46 &  0.75 & 0.73 & 0.77 & 0.78 & 0.71 & 0.91 \\
46--50 &  0.77 & 0.75 & 0.79 & 0.79 & 0.73 & 0.92 \\
50--55 &  0.79 & 0.77 & 0.81 & 0.81 & 0.75 & 0.93 \\
55--60 &  0.81 & 0.79 & 0.83 & 0.83 & 0.77 & 0.94 \\
60--70 &  0.83 & 0.81 & 0.85 & 0.85 & 0.79 & 0.95 \\
70--100 &  0.88 & 0.86 & 0.90 & 0.89 & 0.84 & 0.98 \\
\hline
\end{tabular}
\end{table}

    }
\cleardoublepage \section{The CMS Collaboration \label{app:collab}}\begin{sloppypar}\hyphenpenalty=5000\widowpenalty=500\clubpenalty=5000\textbf{Yerevan Physics Institute,  Yerevan,  Armenia}\\*[0pt]
V.~Khachatryan, A.M.~Sirunyan, A.~Tumasyan
\vskip\cmsinstskip
\textbf{Institut f\"{u}r Hochenergiephysik der OeAW,  Wien,  Austria}\\*[0pt]
W.~Adam, T.~Bergauer, M.~Dragicevic, J.~Er\"{o}, M.~Friedl, R.~Fr\"{u}hwirth\cmsAuthorMark{1}, V.M.~Ghete, C.~Hartl, N.~H\"{o}rmann, J.~Hrubec, M.~Jeitler\cmsAuthorMark{1}, W.~Kiesenhofer, V.~Kn\"{u}nz, M.~Krammer\cmsAuthorMark{1}, I.~Kr\"{a}tschmer, D.~Liko, I.~Mikulec, D.~Rabady\cmsAuthorMark{2}, B.~Rahbaran, H.~Rohringer, R.~Sch\"{o}fbeck, J.~Strauss, W.~Treberer-Treberspurg, W.~Waltenberger, C.-E.~Wulz\cmsAuthorMark{1}
% ...
\vskip\cmsinstskip
% ...
1:~~Also at Vienna University of Technology, Vienna, Austria\\
2:~~Also at CERN, European Organization for Nuclear Research, Geneva, Switzerland\\
% ...

\end{sloppypar}
\end{document}